\documentclass[amssymb, amsmath,nofootinbib, aps, pra, superscriptaddress]{revtex4-2}
\usepackage{graphicx}
\usepackage{verbatim}
\usepackage{xcolor}
\usepackage{esint}
\usepackage{dsfont}
\usepackage[hidelinks]{hyperref}
\begin{document}
\title{Local Scale Invariance in Quantum Theory: A Non-Hermitian Pilot-Wave Formulation}
\author{Indrajit Sen}
\email{isen@ggc.edu}
\affiliation{Department of Physics and Pre-Engineering, Georgia Gwinnett College\\
1000 University Center Lane, Lawrenceville, GA, 30043, USA}
\affiliation{Institute for Quantum Studies, Chapman University\\
One University Drive, Orange, CA, 92866, USA}
\author{Matthew Leifer}
\affiliation{Institute for Quantum Studies, Chapman University\\
One University Drive, Orange, CA, 92866, USA}
\affiliation{Schmid College of Science and Technology, Chapman University\\
One University Drive, Orange, CA, 92866, USA}

\begin{abstract}
We show that Weyl's abandoned idea of local scale invariance has a natural realization at the quantum level in pilot-wave (de Broglie-Bohm) theory. We obtain the Weyl covariant derivative by complexifying the electromagnetic gauge coupling parameter. The resultant non-hermiticity has a natural interpretation in terms of local scale invariance in pilot-wave theory. The conserved current density is modified from $|\psi|^2$ to the local scale invariant, trajectory-dependent ratio $|\psi|^2/ \mathds 1^2[\mathcal C]$, where $\mathds 1[\mathcal C]$ is a scale factor that depends on the pilot-wave trajectory $\mathcal C$ in configuration space. All physical predictions are local scale invariant, even in the presence of mass terms. Our approach is general, and we implement it for the Schrödinger and Pauli equations, and for the Dirac equation in curved spacetime, each coupled to an external electromagnetic field. We also implement it in quantum field theory for the case of a quantized axion field interacting with a quantized electromagnetic field. We discuss the equilibrium probability density and show that the corresponding trajectories are unique. Our results provide a pivotal understanding of local scale invariance in quantum theory.
\end{abstract}

\maketitle

\section{Introduction}
The tension between the two pillars of modern physics, Einstein's general theory of relativity and orthodox quantum theory, lies at the root of numerous research programs in fundamental physics. The apparently intractable nature of this tension suggests that the foundations of both the theories be freely investigated. However, most quantum gravity based approaches to resolving the tension do not question orthodox quantum theory, with the result that the conceptual problems inherent in the latter are carried over. For example, this can be shown to lead to the problem of unitarity in canonical non-perturbative quantum gravity \cite{sentarity}. On the other other hand, most quantum foundations based approaches do not question Einstein's general theory of relativity. This has led to an acute focus on notions of locality natural in general relativity but problematic at the quantum level (such as Bell locality \cite{bell64, bell}). This focus has also contributed to the neglect of notions of locality that are absent in both theories but may be desirable in a future unified theory.\\

A more robust approach would be to abandon both Einstein's general theory of relativity and orthodox quantum theory as fundamental and explore a new starting point. Both theories were challenged early on by alternatives, most of which have been sidelined to history. However, two of them have continued to evoke fascination from the margins. \\

The first alternative is Weyl's gauge theory \cite{weyl18, weyl18a, weyl19, weyl22}, proposed in 1918 as a unified field theory of gravity and electromagnetism. Weyl observed that Riemannian manifold is not purely infinitesimal as vector magnitudes, unlike vector directions, can be compared across distant points on the manifold. Upon removing this inconsistency by introducing his concept of gauge, Weyl obtained a local scale invariant\footnote{Also known as Weyl invariant in the literature.} generalization of general relativity that naturally incorporates electromagnetism. The exceptional beauty of the theory suggested to Weyl its physical realization: ``It would be remarkable if in nature there was realized instead an illogical quasi-infinitesimal geometry...'' \cite{weyl18}. However, the theory was quickly abandoned due to Einstein's criticism that it predicted history-dependent spectral frequencies, contradicting empirical observation \cite{einstein18}. In the late 1920s, the gauge idea was modified to fit orthodox quantum theory \cite{fuck, london27, vizgin, gawn}, leading to local phase invariance (local U(1)) symmetry, and in this modified form survives as a cornerstone of the standard model.\\

Later developments in particle physics and cosmology have led to a resurgence of interest in Weyl's original idea of local scale invariance, which has been linked to a wide variety of topics in fundamental physics: dark matter \cite{darkweyl}, dark energy \cite{shaposhni}, neutrino option \cite{weylneutrino}, inflation \cite{weylflation}, naturalness problem in particle physics \cite{bardeen}, conformal gravity \cite{weylcg} etc. (for an excellent survey, see \cite{scholz18}) Therefore, although neither general relativity nor orthodox quantum theory possesses local scale invariance, it is potentially desirable in a future unified theory as a fundamental symmetry.\\

The second alternative is pilot-wave theory\footnote{Also known as de Broglie-Bohm theory, Bohmian mechanics or the Causal interpretation in the literature.} (PWT) \cite{broglie27}, originally presented by de Broglie at the 1927 Solvay conference. Abandoned soon thereafter (see de Broglie's recollections in \cite{broy}, also see \cite{solventini}), the theory was rediscovered by Bohm \cite{bohm1, bohm2, bohmbook2}, who developed it further and clarified key aspects of its radical inner structure. The theory resolves in a simple manner (from its internal perspective) all the apparent quantum paradoxes, such as fundamental indeterminism, measurement problem \cite{genphqm}, Schrödinger's cat paradox \cite{catpaper}, Wigner's friend paradox \cite{wigfriend}, and their later versions. Further, it is compatible with all the no-go theorems to date, such as Bell's theorem \cite{bell64}, Kochen-Specker theorem \cite{kochi}, and the PBR theorem \cite{pbr}. The theory can be applied to quantized fields and in quantum gravity. However, discussions of PWT in the literature are often mired in misunderstandings \cite{valentrouble} (for a clear introduction to the theory, see \cite{bohm1, bohm2, bohmbook2, hollandbook}). \\

Bohm emphasized that PWT contains within itself the seeds for further developments that are impossible in orthodox quantum theory \cite{bohm1, bohm2, bohmbook2}. In such further developments, PWT can become empirically distinguishable from orthodox quantum theory.\\

In this article, we show that Weyl's gauge theory and PWT are deeply interconnected. A hint of this connection can be discerned from Bohm's observation that ``the quantum potential depends only on the form and not on the amplitude of the quantum field'' \cite[p.~27]{bohmbook2} -- in other words, the quantum potential is global scale invariant. We show that the Weyl covariant derivative is naturally realized in PWT's conceptual structure, thereby promoting the global scale invariance of the quantum potential to local scale invariance. Further, the conserved current density is generalized from the Born rule density to a local scale invariant ratio that depends on the pilot-wave trajectory (particle trajectories in particle quantization, field trajectories in field quantization). That is, the current density at a configuration space point is rendered a functional of the pilot-wave trajectory passing through that point. We show that our approach is general and apply it to a variety of scenarios, incorporating multiple particles, spin, special relativity and field quantization. The resulting theory makes trajectory-dependent empirical predictions and is therefore, in principle, empirically distinguishable from orthodox quantum theory and other quantum formulations. Lastly, we find that our theory illuminates the connections between Hermiticity, canonical commutators and gauge invariance. We explore the experimental predictions of our theory and use it to analyse the Weyl-Einstein debate on the second-clock effect \cite{weyl18, einstein18, vizgin, ryckmanbook} in a subsequent paper \cite{sen26b}. \\

The article is structured as follows. We first motivate our work in a general setting by pointing out connections between canonical commutators, Hermiticity and local scale invariance in section \ref{motiv}. We then show how to implement local scale invariance in non-Hermitian PWT for the case of a single non-relativistic Schrödinger particle coupled to an external electromagnetic field in section \ref{nonrel}. The calculations are presented in full in \ref{nonrel} to clearly illustrate our general approach; for subsequent sections, the detailed calculations are given in the appendix. We treat the case of multiple Schrödinger particles, non-relativistic spin-1/2 particle (Pauli equation), and relativistic spin-1/2 particle (Dirac equation in curved spacetime), each coupled to an external electromagnetic field, in sections \ref{multiple}, \ref{pauli} and \ref{dirac} respectively. We show the applicability of our approach in quantum field theory by treating the case of a quantized axion field interacting with a quantized electromagnetic field in \ref{axion}. We discuss the notion of equilibrium and the uniqueness of the guidance equation in our theory in \ref{prob}. We discuss our results in \ref{disc}.

\section{Canonical commutators, non-Hermiticity and local scale invariance} \label{motiv}
The generalization of the Poisson bracket to the canonical commutator may be considered a fundamental step in quantization that precedes other elements of orthodox quantum theory, such as Hilbert space, unitarity, and linearity \cite{dirac25, diracbook}. We begin our discussion at this step by considering the canonical commutation relation 
\begin{align}
    [\hat x_i, \hat p_j] = i\hbar\delta_{ij} \label{comm}
\end{align}
where $\hat x$ ($\hat p$) is the position (canonical momentum) operator, $i, j \in \{1, 2, 3\}$ label the 3 spatial directions and $\delta$ is the Kronecker-delta function. In the position representation $\hat x_i \to  x_i$, the general form of $\hat p_i$ that satisfies (\ref{comm}) is
\begin{align}
\hat{\vec p} \to -i\hbar \vec \nabla  + \vec f(\vec x,t) \label{mom}
\end{align}
where $\vec f(\vec x, t)$ is an arbitrary vectorial function of position and time. \\

The freedom to choose $\vec f(\vec x, t)$ is intimately connected to electromagnetic gauge freedom. Consider the Schrödinger equation for a spinless particle in external electromagnetic field. The kinetic energy term in the Hamiltonian is $(-i\hbar \vec \nabla - e\vec A)^2/2m$, where $\hat{\vec p} = -i\hbar \vec \nabla$, $e$ ($m$) is the charge (mass) of the particle and $\vec A$ is the electromagnetic vector potential. The kinetic energy term changes from $(\hat{\vec{p}} - e\vec A)^2/2m \to (\hat{\vec{p}} - e\vec A - e\vec \nabla \lambda)^2/2m$ upon the gauge transformation $\vec A \to \vec A + \vec \nabla \lambda$. The same result is achieved if, instead of performing a gauge transformation of the vector potential, we change the $\hat{\vec p}$ representation from $-i\hbar \vec \nabla \to -i\hbar \vec \nabla -e\vec \nabla \lambda$ by setting $\vec f(\vec x,t) = -e\vec \nabla \lambda$ in equation (\ref{mom}). Therefore, the gauge transformation $\vec A \to \vec A + \vec \nabla \lambda$ and the canonical momentum operator transformation $-i\hbar \vec \nabla \to -i\hbar \vec \nabla -e\vec \nabla \lambda$ are equivalent here. \\

Note that, although $\vec f(\vec x, t) = -e\vec \nabla \lambda$ is purely real, a complex $\vec f(\vec x, t)$ in (\ref{mom}) also satisfies the canonical commutation relation (\ref{comm}). This implies that the full solution of the canonical momentum operator (\ref{mom}) is not utilized in orthodox quantum theory. This is also suggested from the dual roles played by $e$ in the kinetic energy term $(-i\hbar \vec \nabla - e\vec A)^2/2m$. The variable $e$ simultaneously represents $a)$ the charge of the particle, and $b)$ the coupling parameter for interaction with the electromagnetic gauge field. The particle charge must clearly be real, but the coupling parameter, which determines the effect of $A$ on the complex quantum state, can in principle be complex. This suggests that we can separate these two conceptually distinct roles by replacing $e \to e_C \equiv e + i e_I$, where the imaginary (real) part of the coupling parameter $e_C$ is $e_I$ ($e$). This leads to the most general form of the canonical momentum operator in (\ref{mom}) allowed by the commutation relation (\ref{comm}).\\

The replacement $e \to e_C$ leads to a non-Hermitian kinetic energy term $(-i\hbar \vec \nabla - e_C\vec A)^2/2m$, which contradicts a basic postulate of orthodox quantum theory that the Hamiltonian for a closed system is Hermitian\footnote{Open quantum systems can be modelled using effective non-Hermitian hamiltonians \cite{open}, which are irrelevant to the discussion here. The full, closed system containing the environment evolves via a Hermitian hamiltonian in the orthodox formulation.}. The Hermiticity postulate is often justified by appealing to the reality of observed eigenvalues, although the latter does not actually imply Hermiticity \cite{bend98, benderbook}. From a foundational perspective, we know that the route from postulates to empirical observations is theory dependent. Therefore, the Hermiticity postulate is neither a mathematical nor a conceptual necessity in an alternative quantum formulation, such as PWT. \\

Admitting the non-Hermitian representation of the canonical momentum operator opens up the possibility of implementing, at the quantum level, Weyl's original idea of gauge invariance as local scale invariance. This is because the gauge covariant derivative originally proposed by Weyl (known as Weyl covariant derivative in the literature) is of the form $\vec \nabla - \omega \vec A$, where $\omega$ is a real number known as the Weyl weight, which is determined by the scale transformation properties of the field on which the covariant derivative is acting \cite{weyl18}. The Weyl weight controls non-integrable scale effects \cite{weyl18}. Comparing this to the non-Hermitian covariant derivative $-i\hbar \vec \nabla - e_C\vec A$, we can see that the imaginary component of the gauge coupling parameter $e_I$ implements the Weyl covariant derivative. The derivative here acts on the wavefunction $\psi = Re^{iS/\hbar}$, which can be considered to be a field of vectors in configuration space, with the local vector magnitude given by the amplitude $R$ and the local vector direction given by the phase $S$. As we show in subsequent sections, the presence of a non-zero $e_I$ leads to local scale invariance joining local phase invariance as a symmetry of the theory in non-Hermitian PWT.

\section{Spinless non-relativistic particle} \label{nonrel}
We illustrate our general approach here by discussing the case of a spinless, non-relativistic particle coupled to an external electromagnetic field. We use the 4-vector notations $x^\mu \equiv (ct, \vec{x})$, $A^\mu \equiv (\phi, \vec A)$, $\mu = \{0, 1,2, 3\}$ and adopt the mostly-negative Minkowski metric $\eta_{\mu \nu} \equiv \text{diag}(+, -, -, -)$. The Schrödinger equation for the system is given in the position representation by
\begin{align}
    \big [\frac{(-i\hbar\vec{\nabla} -e_C\vec{A}/c) \cdot (-i\hbar\vec{\nabla} -e_C\vec{A}/c)}{2m} + e_C\phi\big] \psi(\vec{x},t) = i \hbar \frac{\partial \psi(\vec{x},t)}{\partial t} \label{schr}
\end{align}
where $\vec{A}(\vec{x},t)$ and $\phi(\vec{x},t)$ are the electromagnetic vector and scalar potentials respectively. We first briefly review the pilot-wave analysis of the system for the Hermitian case $e_C = e$.\\

Using the ansatz $\psi(\vec{x},t) = R(\vec{x},t) e^{i S(\vec{x},t)/\hbar}$ and $e_C = e$ in (\ref{schr}), we can obtain the following two real equations
\begin{align}
    \frac{\partial S}{\partial t} + \frac{(\vec{\nabla}S - e \vec{A}/c)^2}{2m} + e\phi - \frac{\hbar^2}{2m} \frac{\nabla^2 R}{R} = 0 \label{hj0}\\
    \frac{\partial R^2}{\partial t} + \vec{\nabla}\cdot (R^2 \frac{\vec{\nabla}S - e \vec{A}/c}{m}) = 0 \label{con0}
\end{align}
Equation (\ref{hj0}) is the quantum Hamilton-Jacobi equation. It implies that the particle piloted by the quantum state has kinetic momentum $(\vec{\nabla}S - e\vec A)$ and is subject to the quantum potential $-(\hbar^2/2m) \nabla^2 R/R$ in addition to the classical potential $e\phi$. Equation (\ref{con0}) is the quantum continuity equation and describes a conserved current associated with the evolution of the quantum state. It defines the quantum equilibrium density $R^2$ for an ensemble of particles piloted by the quantum state. \\

Suppose that $e_C$ is complex, such that $e_C = e + i e_I$. The electric charge $e$ is then the real component of the coupling parameter $e_C$. In addition, there is an imaginary component $e_I$ of the coupling parameter. The form of the Schrödinger equation (\ref{schr}) remains unchanged, but $e_I$ leads to non-Hermitian terms of the form $i (\vec A \cdot \vec p + \vec p \cdot \vec A)/2mc$, $i A^2/2mc^2$ and $i \phi$ in the hamiltonian, where $A^2 \equiv \vec A \cdot \vec A$. Let us investigate the effect of $e_I$ on gauge invariance.\\

Consider a general gauge transformation $A^\mu \to A^\mu - \partial^\mu \lambda$, where $\lambda(x^\mu)$ is an arbitrary real function of $x^\mu$. The Schrödinger equation (\ref{schr}) transforms to 
\begin{align}
    \big [\frac{(-i\hbar\vec{\nabla} -e_C\vec{A}/c - e_C\vec{\nabla}\lambda/c) \cdot (-i\hbar\vec{\nabla} -e_C\vec{A}/c - e_C\vec{\nabla}\lambda/c)}{2m} + e_C\phi - e_C\frac{\partial \lambda}{\partial ct}\big] \psi'(\vec{x},t) = i \hbar \frac{\partial \psi'(\vec{x},t)}{\partial t} \label{schrg}
\end{align}
where $\psi'$ is the transformed wavefunction. It is straightforward to show that equation (\ref{schrg}) reduces to (\ref{schr}) under the transformation 
\begin{align}
    \psi \to \psi' = \psi e^{i\frac{e_C\lambda}{\hbar c}} \label{gta}
\end{align}
Interestingly, then, the gauge invariance of the equation is unaffected by the introduction of $e_I$. We note that as $e_C$ is complex, the wavefunction undergoes a local \textit{scale} as well as a local phase transformation in equation (\ref{gta}).\\

Let us investigate this system from a pilot-wave perspective. Using the polar decomposition of the quantum state, we can obtain the following two real equations
\begin{align}
 \frac{\partial S}{\partial t} + \frac{(\vec{\nabla}S - e \vec{A}/c)^2}{2m} + e\phi -\frac{\hbar^2}{2m}\bigg(\frac{\nabla^2 R}{R} +\frac{e_I^2 A^2}{\hbar ^2 c^2} + \frac{e_I}{\hbar c}\big( \vec{\nabla}\cdot \vec{A}+2\frac{\vec{A}\cdot \vec{\nabla}R}{R}\big) \bigg) = 0 \label{hja2} \\
     \frac{\partial R^2}{\partial t} + \vec{\nabla}\cdot (R^2 \frac{\vec{\nabla}S - e \vec{A}/c}{m}) = 2R^2\bigg(-\frac{e_I}{\hbar c} \vec{A}\cdot \frac{(\vec{\nabla}S - e \vec{A}/c)}{m}  + \frac{e_I}{\hbar} \phi\bigg) \label{cona2}
\end{align}
Equation (\ref{hja2}) appears to retain its form as the quantum Hamilton-Jacobi equation. It implies that the particle kinetic momentum continues to be given by $(\vec{\nabla}S - e \vec{A}/c)$, as in equation (\ref{hj0}). The quantum potential, however, is modified by additional $e_I$-dependent terms. Equation (\ref{cona2}) appears to imply, on comparison with (\ref{con0}), that the non-Hermiticity leads to sink/source terms on the right-hand side of the continuity equation. Let us investigate both the equations.\\

\subsection{Quantum Continuity equation} \label{schrcon}
We can rewrite equation (\ref{cona2}) as
\begin{align}
    D_t R^2 + \vec D \cdot (R^2 \frac{\vec \nabla S - e \vec A/c}{m}) = 0 \label{conya2}
\end{align}
where the ordinary derivatives have been replaced by the Weyl covariant derivatives
\begin{align}
    \partial_\mu \to D_\mu \equiv (\partial_\mu - \frac{\omega e_I}{\hbar c} A_\mu) \label{cove}
\end{align}
acting on $R^2$, which has Weyl weight $\omega = 2$, determined by the transformation $R^2 \to R'^2 = R^2 e^{-\frac{2e_I \lambda(x)}{\hbar c}}$ under a gauge transformation. Clearly, the imaginary component of the coupling parameter $e_I$ defines the Weyl-covariant derivative (\ref{cove}). Equation (\ref{conya2}) implies that $R^2$ is conserved with respect to the covariant derivatives but not with respect to ordinary derivatives.\\ 

The particle velocity 
\begin{align}
    \vec v = \frac{\vec j}{j_0} = \frac{\vec{\nabla}S - e \vec{A}/c}{m} \label{noguy}
\end{align}
is consistent with the kinetic momentum obtained from the quantum Hamilton Jacobi equation (\ref{hja2}). Let us check the gauge invariance of equation (\ref{noguy}). We know that, upon a gauge transformation $A^\mu \to A^\mu - \partial^\mu \lambda $, the quantum state changes from $\psi \to \psi' = \psi e^{i\frac{e_C\lambda}{\hbar c}}$. It follows that the phase changes from $S \to S' = S + e\lambda/c$, so that equation (\ref{noguy}) is gauge invariant.\\

Let us suppose that the density conserved with respect to the ordinary derivatives is given by a gauge-invariant ratio $R^2/\Omega$, where $\Omega$ is a weight factor to be determined. Equation (\ref{cona2}) then implies that
\begin{align}
    \frac{\partial}{\partial t}(\frac{R^2}{\Omega})+ \vec{\nabla}\cdot (\frac{R^2}{\Omega} \frac{\vec{\nabla}S - e \vec{A}/c}{m}) = R^2\bigg(-2\frac{e_I}{\Omega \hbar c} \vec{A}\cdot \frac{(\vec{\nabla}S - e \vec{A}/c)}{m} - \frac{(\vec{\nabla}S - e \vec{A}/c)}{m\Omega^2 }\cdot \vec{\nabla}\Omega + \frac{2e_I}{\Omega \hbar} \phi - \frac{1}{\Omega^2}\frac{\partial \Omega}{\partial t} \bigg)
\end{align}
For $R^2/\Omega$ to be conserved, we must have
\begin{align}
   -2\frac{e_I}{\Omega \hbar c} \vec{A}\cdot \frac{(\vec{\nabla}S - e \vec{A}/c)}{m} - \frac{(\vec{\nabla}S - e \vec{A}/c)}{m\Omega^2 }\cdot \vec{\nabla}\Omega + \frac{2e_I}{\Omega \hbar} \phi - \frac{1}{\Omega^2}\frac{\partial \Omega}{\partial t}  = 0 \label{constr}
\end{align}
which can be rewritten as
\begin{align}
  v^{\mu} D_\mu \Omega = 0 \label{gconstr}
\end{align}
where the Weyl covariant derivative $D_\mu$, defined by equation (\ref{cove}), acts on $\Omega$, which has Weyl weight $\omega = 2$ as the ratio $R^2/\Omega$ is gauge invariant. Further, $v^0 = dx^0/dt = c$ and $\vec v$ is defined by the guidance equation \eqref{noguy}. Equation (\ref{gconstr}) implies that $\Omega$ is parallel-transported in Weyl geometry along the particle trajectories. It can also be expressed as 
\begin{align}
    \frac{d\Omega}{\Omega} = \frac{2e_I}{\hbar c} A_\mu dx^\mu \label{beauty}
\end{align}
where $d\Omega \equiv \partial_\mu \Omega dx^\mu$ is the change in $\Omega$ along the particle trajectory. Equation \eqref{beauty} has the path-dependent solution $\Omega = \Omega_0(x_0) e^{-\frac{2e_I}{\hbar c}(\int^{\vec x}_{\mathcal{C}} \vec{A}\cdot d\vec{x}' - \int^t_\mathcal{C} \phi c dt')} = \Omega_0(x_0) e^{\frac{2e_I}{\hbar c}\int^x_\mathcal{C} A^{\mu'} dx_{\mu'}}$, where $\int^x_\mathcal{C} A^{\mu'} dx_{\mu'} \equiv \int_\mathcal{C}^t \phi c dt' - \int^{\vec x}_\mathcal{C} \vec{A}\cdot d\vec{x}'$ is a path-dependent line integral\footnote{The integral is path dependent because its mixed partial derivatives do not commute in general, as the electromagnetic field tensor $F_{\mu \nu} = \partial_\mu A_\nu - \partial_\nu A_\mu \neq 0$ in general. The integral is path independent in cases $F_{\mu \nu} = 0$. Note that the path-dependent line integral should not be confused with Feynman's path-integral.} along the space-time curve $\mathcal{C} = \{ (ct'(\lambda), \vec x'(\lambda))| \lambda \in [0,1] \}$ determined by the guidance equation (\ref{noguy}). Here $\Omega_0(x_0)$ is the value of $\Omega$ at $x_0 \equiv (ct_0, \vec x_0)$, which is the initial space-time point of the curve $\mathcal C$. \\

To determine $\Omega_0$, we impose the condition that our theory reduce to orthodox quantum theory when $e_I$ is set to zero. This implies that the initial gauge-invariant current density is equal to the Born-rule density, and that any deviation from orthodox quantum theory must arise from the effect of $e_I$ on the subsequent evolution of the system. We thus have the initial condition:
\begin{align}
\frac{R^2(x_0)}{\Omega_0(x_0)} = \tilde R^2(x_0) \label{initialize}
\end{align}
where $\tilde R(x_0)$ is the gauge-invariant amplitude in orthodox quantum theory at the initial time $t= t_0$ (state preparation). Equation \eqref{initialize} implies that if the initial amplitude satisfies $R^2(x_0) = \tilde R^2(x_0) e^{-\frac{2e_I \lambda(x_0)}{\hbar c}}$ in a particular gauge, then $\Omega_0(x_0) = 1 \cdot e^{-\frac{2e_I \lambda(x_0)}{\hbar c}}$ in that gauge. In the gauge where the initial gauge-dependent amplitude satisfies $R(x_0) = \tilde R(x_0)$, $\Omega_0(x_0) = 1$. If $e_I = 0$, then $\Omega_0(x_0) = 1$ is gauge invariant and equation \eqref{beauty} implies that $\Omega(x) = 1$ for all $t \geq  t_0$. The gauge-invariant current density $R^2(x)/ \Omega$ thereby becomes equal to $\tilde R^2(x)$ at all times if $e_I =0$, and orthodox quantum theory is recovered. Further, equation \eqref{beauty} implies that if the condition \eqref{initialize} is not satisfied, then $R^2(x)/ \Omega \neq \tilde R^2(x)$ for all $t \geq t_0$ even if $e_I = 0$. Henceforth, we work in the gauge\footnote{We can always find this gauge if $R(x_0) \neq \tilde R(x_0)$ in a chosen gauge by setting $R(x_0)e^{-\frac{e_I \lambda(x_0)}{\hbar c}} = \tilde R(x_0)$ and solving for $\lambda$.} which satisfies $R(x_0) = \tilde R(x_0)$ for convenience, it being understood that all physical results are gauge invariant. In this gauge $\Omega_0(x_0) = 1$, which implies that
\begin{align}
    \Omega = \mathds 1^2[\mathcal C] \label{solulu}
\end{align}
where 
\begin{align}
    \mathds 1[\mathcal C] = e^{\frac{e_I}{\hbar c}\int^x_\mathcal{C} A^{\mu'} dx_{\mu'}} \label{scale}
\end{align}
is the parallel-transported scale from space-time point $x_0 \to x$ along the pilot-wave trajectory $\mathcal C$, specified by the guidance equation \eqref{noguy}. The current density is given by the gauge-invariant ratio $R^2/\mathds 1^2[\mathcal C]$. This neatly implements Weyl's original idea of gauge wherein physical magnitudes are defined with respect to local scales -- that is, they are local ratios -- and are local scale invariant. Further, physical magnitudes defined at distant points on the manifold cannot be directly compared (without parallel transport). The physical magnitude of the initial amplitude $R(x_0)$, in the gauge which satisfies $R(x_0) = \tilde R(x_0)$, is defined with respect to a local scale of magnitude 1 at each spatial point $\vec x_0$ at $t_0$. Equivalently, the physical magnitude of the initial amplitude $R(x_0)$, in the gauge which satisfies $R(x_0) = \tilde R(x_0) e^{-\frac{e_I \lambda(x_0)}{\hbar c}}$, is defined with respect to a local scale of magnitude $1 \cdot e^{-\frac{e_I \lambda(x_0)}{\hbar c}}$ at each spatial point $\vec x_0$ at $t_0$. The local scale at $x_0$ is then parallel transported, along the pilot-wave trajectory $\mathcal C$, to $x$. In the gauge which satisfies $R(x_0) = \tilde R(x_0)$ $\big (R(x_0) = \tilde R(x_0) e^{-\frac{e_I \lambda(x_0)}{\hbar c}}\big )$, the local scale at $x$ is given by $\mathds 1[C]$ $\big (\mathds 1[C]\cdot e^{-\frac{e_I \lambda(x)}{\hbar c}}\big )$. The physical magnitude of the final amplitude is the local ratio $R(x)/\mathds 1[\mathcal C]$ at $x$. Note that the ratio $R(x)/\mathds 1[\mathcal C]$ is single-valued at $x$ as the pilot-wave velocity field (\ref{noguy}) is defined on configuration space, which implies that each space-time point in the support of $\psi(\vec x, t)$ is mapped to a single particle trajectory that starts at time $t_0$ and ends at that point. \\

Finally, the continuity equation (\ref{cona2}) becomes
\begin{align}
    \frac{\partial}{\partial t}(\frac{R^2}{ \mathds 1^2}) + \vec{\nabla}\cdot (\frac{R^2}{ \mathds 1^2} \frac{\vec{\nabla}S - e \vec{A}/c}{m}) = 0 \label{finale}
\end{align}
where we have omitted the trajectory symbol $\mathcal C$ in $\mathds 1^2$ for brevity. Equation (\ref{finale}) reduces to the usual continuity equation whenever the effect of the local scale change is negligible. That is, if $e_I = 0 $ then $\mathds 1^2 = 1$ and $R^2 = \tilde R^2$, where $\tilde R$ is the amplitude in orthodox quantum theory.\\

Lastly, we can normalize the quantum state by the condition 
\begin{align}
    \int \frac{R^2}{ \mathds 1^2} d^3\vec x = 1
\end{align}
which is preserved in time by (\ref{finale}). 

\subsection{Quantum Hamilton-Jacobi equation}
We can express the modified quantum potential term in equation (\ref{hja2}) concisely as 
\begin{align}
    \frac{\nabla^2 R}{R} +\frac{e_I^2 A^2}{\hbar ^2 c^2} + \frac{e_I}{\hbar c}\big(  \vec{\nabla}\cdot \vec{A}+2\frac{\vec{A}\cdot \vec{\nabla}R}{R}\big) = \frac{ D^2 R }{R} \label{qrockpot}
\end{align}
where $D^2 \equiv \vec D \cdot \vec D$ and $\vec D = (1 + e_I \vec A/\hbar c)$ is the spatial component of the Weyl covariant derivative, defined by \eqref{cove}, acting on $R$, which has Weyl weight $\omega = 1$. The Weyl weight here is determined by the transformation $R \to R' = R e^{-\frac{e_I \lambda(x)}{\hbar c}}$ under a gauge transformation. It is straightforward to verify that $D^2 R/R$ is local scale invariant.
\begin{figure}
    \centering
    \includegraphics[width=\linewidth]{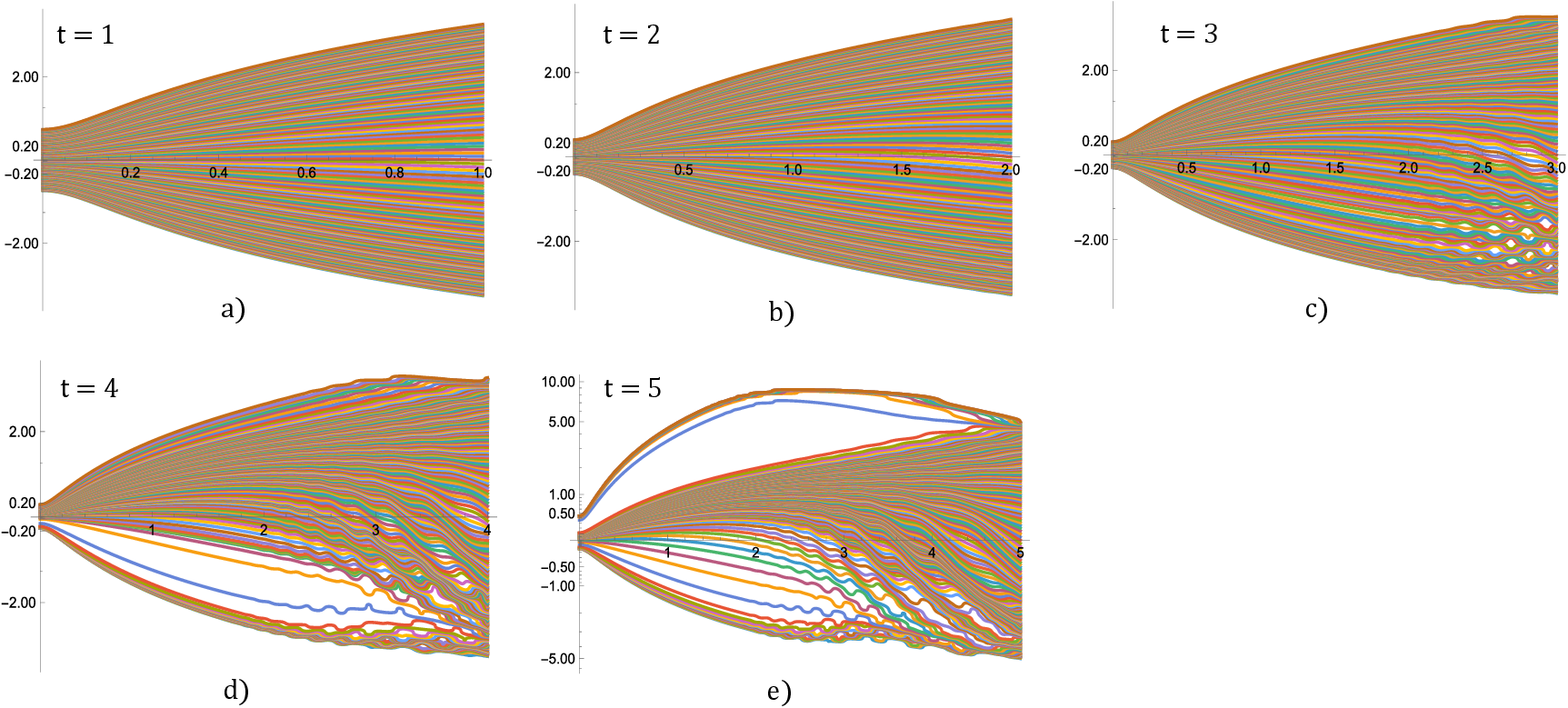}
    \caption{Computed particle trajectories for the non-Hermitian potential $V(x) = i \sin x$. The trajectories crossing $x \in [-5,5]$ in $0.05$ equally-spaced increments at times $t = 1, 2...5$ s are shown. The horizontal (vertical) axis corresponds to $t$ ($x$). The trajectories are required to compute the scale factor $\mathds 1[\mathcal C]$ in the conserved current density $|\psi|^2/\mathds 1^2[\mathcal C]$. The vertical axis is logarithmically scaled.}
    \label{ftrajectories}
\end{figure}
Equation (\ref{hja2}) can then be expressed as
\begin{align}
     \frac{\partial S}{\partial t} + \frac{(\vec{\nabla}S - e \vec{A}/c)^2}{2m} + e\phi - \frac{\hbar^2}{2m}\frac{D^2 R }{R} = 0 \label{hja2'}
\end{align}
Clearly, given $e_I$, the real part of the Schrödinger equation continues to function as the quantum Hamilton-Jacobi equation. Equation \eqref{hja2'} also clarifies the complementary roles played by the components of the complex coupling parameter $e_C = e + i e_I$. The real component $e$ modifies the form of the kinetic momentum to make it gauge invariant (local phase invariant), whereas the imaginary component $e_I$ modifies the form of the quantum potential to make it gauge invariant (local scale invariant). It is straightforward to check that the quantum Hamilton-Jacobi equation \eqref{hja2'} is gauge invariant.\\
\begin{figure}
    \centering
    \includegraphics[width=\linewidth]{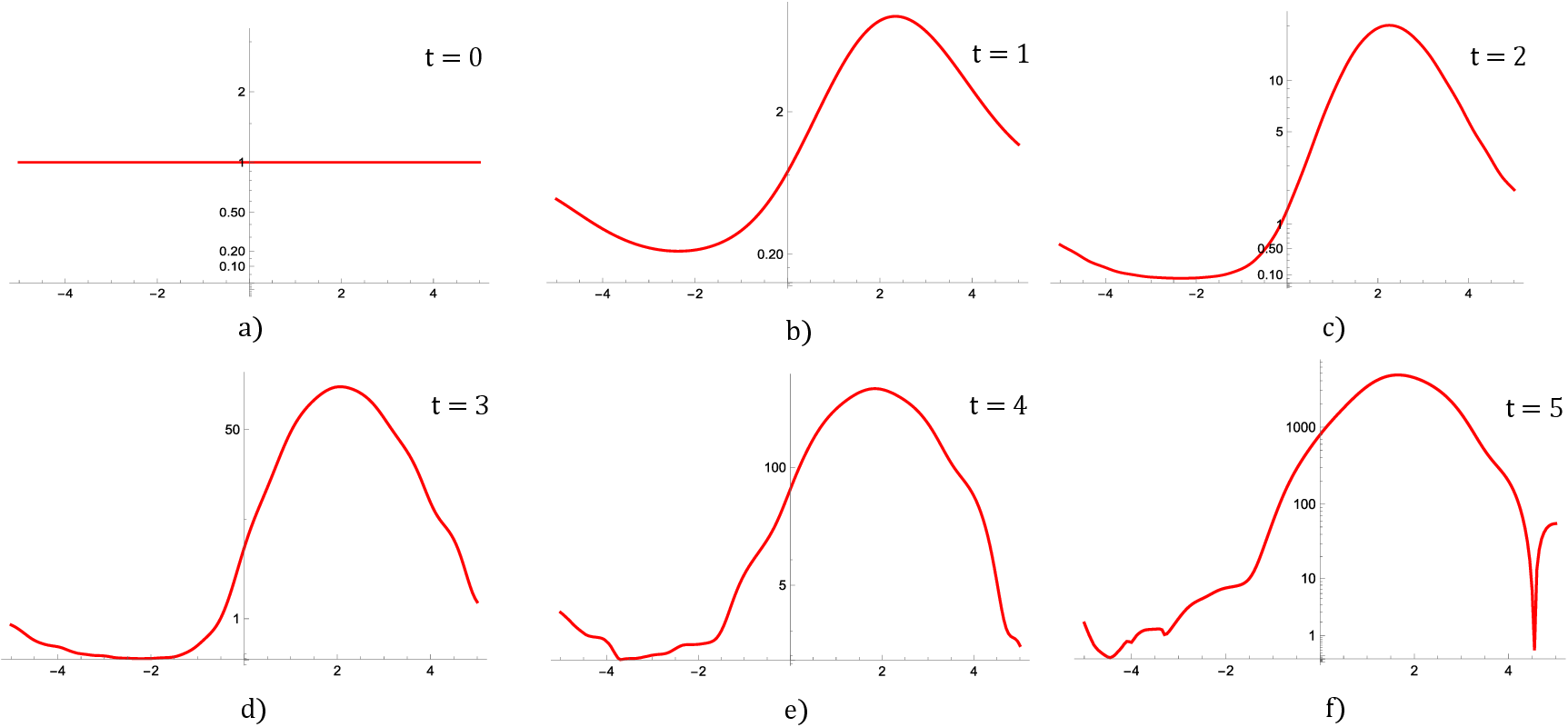}
    \caption{Computed scale factor squared $\mathds 1^2[\mathcal C]$ for the non-Hermitian potential $V(x) = i \sin x$ at times $t = 1, 2...5$ s. The scale factor $\mathds 1[\mathcal C]$ at each point $(x, t)$ is defined as the line integral $\mathds 1[\mathcal C] = e^{\int^{(x,t)}_\mathcal{C} \sin x'(t')\textbf{ } dt'}$ over the particle trajectory from $(x_0, 0)$ to $(x, t)$, where $x_0$ is the particle position at $t=0$ s. The particle trajectories are shown in figure \ref{ftrajectories} and the density $|\psi|^2/\mathds 1^2[\mathcal C]$ is shown in figure \ref{fdensity}. The vertical axis is logarithmically scaled.}
    \label{fscale}
\end{figure}
\subsection{Illustration: particle in $V(x) = i\sin x$ potential}\label{example}
It is useful to consider a simple example to illustrate our approach. Consider a 1D non-relativistic particle in the non-Hermitian, $\mathcal{PT}$-symmetric \cite{benderbook} potential $V(x) = i \sin x $. The potential $V(x)$ can be modelled by setting $A(x) = 0, \phi(x) = \sin x$ and setting $e = 0, e_I = 1$. The Schrödinger equation is $i \partial_t \psi = -\partial_x^2 \psi/2 + i \sin x \psi$, where we have set $m = \hbar = c = 1$ for convenience.\\ 

\begin{figure}
    \centering
    \includegraphics[width=\linewidth]{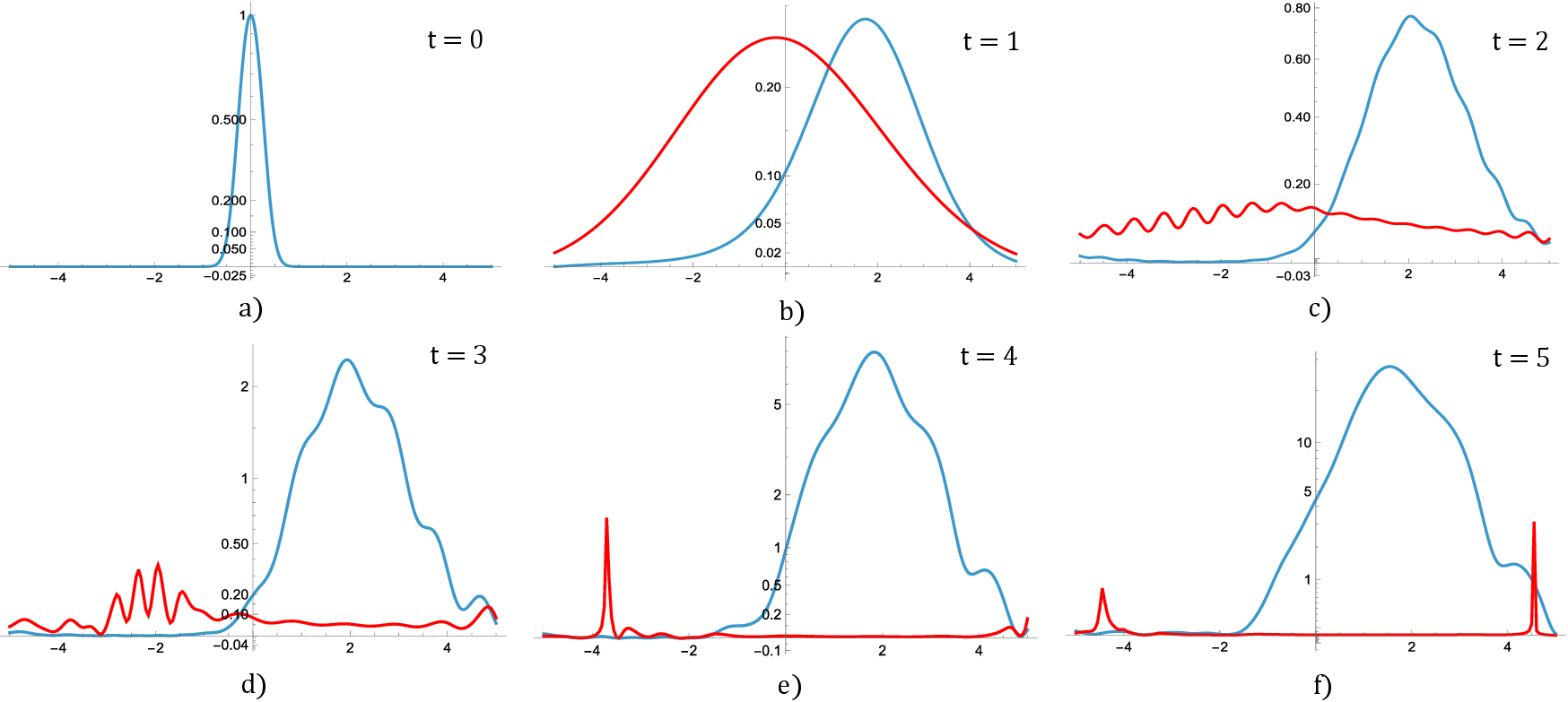}
    \caption{Comparison of the computed densities $|\psi|^2/\mathds 1^2[\mathcal C]$ (\textcolor{red}{red}) and $|\psi|^2$ (\textcolor{blue}{blue}) for the non-Hermitian potential $V(x) = i \sin x$ at times $t = 1, 2...5$ s. The two densities are identical at $t = 0$ but quickly diverge. The density $|\psi|^2/\mathds 1^2[\mathcal C]$ is conserved and remains normalized, whereas $|\psi|^2$ is not conserved and grows large over extended regions. Comparison with figure \ref{fscale} shows that $\mathds 1^2[\mathcal C]$ grows large in the regions where $|\psi|^2$ is large so as to keep the ratio $|\psi|^2/\mathds 1^2[\mathcal C]$ normalized. The vertical axis is logarithmically scaled.}
    \label{fdensity}
\end{figure}

The velocity field (\ref{noguy}) is used to compute the trajectories crossing $x \in [-5,5]$, at uniformly-spaced intervals of $0.05$, at times $t = 0, 1, 2 ...5 $ s. The trajectories are shown in figure \ref{ftrajectories}. The scale factor is computed using (\ref{scale}) as the line integral $\mathds 1[\mathcal C] = e^{\int^{(x,t)}_\mathcal{C} \sin x'(t')\textbf{ } dt'}$ over the trajectories. The scale factor squared $\mathds 1^2[\mathcal C]$ in shown in figure \ref{fscale}. The conserved current density $|\psi|^2/\mathds 1^2[\mathcal C]$ and the unconserved $|\psi|^2$ density are both shown in figure \ref{fdensity}.

\section{Multiple non-relativistic spinless particles} \label{multiple}
Let us extend our approach to multiple particles. The Schrödinger equation for $N$ spinless particles in the position representation is given by
\begin{align}
    \bigg (\sum_{j= 1}^N \big [\frac{(-i\hbar\vec{\nabla}_j -e_{Cj}\vec{A}_j/c) \cdot (-i\hbar\vec{\nabla}_j -e_{Cj}\vec{A}_j/c)}{2m_j} + e_{Cj}\phi_j\big] + V(\vec X,t) \bigg ) \psi(\vec X,t) = i \hbar \frac{\partial \psi(\vec X,t)}{\partial t} \label{schrn}
\end{align}
where $\vec X \equiv (\vec x_1, \vec x_2, \vec x_3...\vec x_N)$, $j \in \{1, 2, 3... N\}$, $\vec \nabla_j \equiv \frac{\partial}{\partial \vec x_j}$, $\vec A_j \equiv \vec A(\vec x_j, t)$, $\phi_j \equiv \phi(\vec x_j, t)$, $V(\vec X, t)$ is the interaction potential between the particles and $e_{Cj} \equiv e_j + i e_{Ij}$ is the gauge coupling parameter of the $j^{th}$ particle.\\

The real part of (\ref{schrn}), analogous to the single-particle case, leads to the quantum Hamilton-Jacobi equation
\begin{align}
    \frac{\partial S}{\partial t} + \sum_{j=1}^N \bigg (\frac{(\vec{\nabla}_jS - e_j \vec{A}_j/c)^2}{2m_j} + e_j\phi_j  -\frac{\hbar^2}{2m_j}\frac{D_j^2 R}{R}\bigg) = 0 \label{hjan}
\end{align}
where $\vec D_j \equiv (\vec \nabla_j + \omega e_I\vec A_j/\hbar c)$ is the $j^{th}$ spatial component of the Weyl covariant derivative acting on $R$, which has Weyl weight $\omega =1$. It is straightforward to check that equation (\ref{hjan}) is gauge invariant. The velocity field is given by 
\begin{align}
    \vec v_j = \frac{\vec \nabla_j S - e_j \vec A_j/c}{m_j},\textbf{ } \forall j \label{nonguy}
\end{align}
It can be shown (see appendix \ref{multipla}) that the continuity equation is 
\begin{align}
     \frac{\partial}{\partial t}(\frac{R^2}{\mathds 1^2}) + &\sum_{j=1}^N\vec \nabla_j \cdot (\frac{R^2}{\mathds 1^2} \frac{\vec \nabla_j S - e_j \vec A_j/c}{m_j}) = 0 \label{conyann}
\end{align}
where $\mathds 1 = \Pi_{j=1}^N e^{\frac{e_{Ij}}{\hbar c}\int^{(ct, \vec{X})}_\mathcal{C} A_{j\mu'} dx^{\mu'}_j}$ is the scale parallel transported from $(\vec X_0, t_0) \to (\vec X, t)$ along the $N$-particle configuration space trajectory $\mathcal{C} = \{ (t'(\lambda), \vec X'(\lambda))| \lambda \in [0,1] \}$, which is determined by the nonlocal velocity field (\ref{nonguy}).

\section{Spin-1/2 non-relativistic particle} \label{pauli}
Let us incorporate spin in our approach by applying it to the case of a spin-1/2 non-relativistic particle\footnote{Note that the applicability of the quantum Hamilton-Jacobi equation for spin-1/2 particles is controversial (see, for example, Bohm's argument in \cite{bohmbook2hj}). We therefore focus only on the guidance equation, which is sufficient to make the theory well-defined.} in external electromagnetic field. The system is described by the Pauli equation, which is given by 
\begin{align}
    \big [\frac{(-i\hbar\vec{\nabla} -e_C\vec{A}/c) \cdot (-i\hbar\vec{\nabla} -e_C\vec{A}/c) \hat I_2  -\hbar e_C \vec B \cdot \hat{\vec \sigma}/c}{2m}  +  e_C\phi \hat I_2 \big] |\psi(\vec x, t)\rangle  = i \hbar \frac{\partial |\psi(\vec x, t)\rangle }{\partial t} \label{paul}
\end{align}
where $|\psi(\vec x, t)\rangle \equiv \psi_+(\vec x, t) |+\rangle_z +  \psi_-(\vec x, t)|-\rangle_z $ is a 2-component spinor (written in the $\hat \sigma_z$ basis), $\hat I_2 $ is the $2 \times 2$ identity matrix and $\hat {\vec \sigma} \equiv (\hat\sigma_x, \hat\sigma_y, \hat\sigma_z)$ are the Pauli matrices. The continuity equation can be shown to be (see appendix \ref{paula})
\begin{align}
     &\frac{\partial}{\partial t}\bigg (\frac{R_+^2 + R_-^2}{\mathds 1^2}\bigg) + \vec{\nabla}\cdot \bigg (\frac{1}{\mathds 1^2} \big(R_+^2 \frac{\vec{\nabla}S_+ - e \vec{A}/c}{m}+R_-^2 \frac{\vec{\nabla}S_- - e \vec{A}/c}{m} +\frac{\vec D \times \vec s}{m} \big )\bigg ) = 0 \label{kunpai}
\end{align}
where $\vec D = \vec \nabla + \frac{\omega e_I}{\hbar c} \vec A$ is the spatial component of the Weyl covariant derivative acting on the local spin density $\vec s \equiv \frac{\hbar}{2} \langle \psi(\vec x, t)|\hat{\vec \sigma}|\psi(\vec x, t) \rangle$, which has Weyl weight $\omega =2$, and $\mathds 1 = e^{\frac{e_I}{\hbar c}\int^x_\mathcal{C} A^{\mu'} dx_{\mu'}}$ is the parallel-transported scale from space-time point $x_0 \to x$ along the particle trajectory $\mathcal C$, defined by the guidance equation
\begin{align}
    \vec v \equiv \frac{\vec j}{j_0} =  \frac{1}{m}\frac{R_+^2 (\vec{\nabla}S_+ - e \vec{A}/c)+R_-^2 (\vec{\nabla}S_- - e \vec{A}/c) +\vec D \times \vec s }{R_+^2 + R_-^2} \label{spinvel}
\end{align}
It is straightforward to check that the velocity field (\ref{spinvel}) is gauge invariant. Note the presence of the spin-term $\vec D \times \vec s$ in (\ref{spinvel}). It is also straightforward to generalize our approach to multiple spin-1/2 particles, similar to section \ref{multiple}, but we do not pursue this here.

\section{Spin-1/2 relativistic particle} \label{dirac}
We generalize here to the relativistic regime by implementing our approach for the case of a Dirac particle in curved spacetime, coupled to an external electromagnetic field. The Dirac equation is given by 
\begin{align}
    \gamma^\mu (i\hbar \nabla_\mu- \hat I_4 \frac{e_C}{c} A_\mu)\psi = mc \psi \label{diro}
\end{align}
where the spacetime covariant derivative $\nabla_\mu \equiv (\hat I_4\partial_\mu + \Gamma_\mu )$, $\Gamma_\mu$ is the spinor connection in curved spacetime and $\hat I_4$ is the $4\times 4$ identity matrix. The quantum state $\psi = \psi^a$ ($a \in \{1, 2,3, 4\}$) is a 4-component spinor and $\gamma^\mu$ are the $4\times 4$ Dirac matrices that satisfy the anti-commutation relation $\{\gamma^\mu, \gamma^\nu\} = 2g ^{\mu\nu}\hat I_4$, where $g^{\mu\nu}$ is the contravariant metric tensor. The continuity equation can be shown to be (see appendix \ref{diraa}) 
\begin{align}
   \partial_\mu (\sqrt{-g}\frac{\overline \psi \gamma^\mu \psi}{\mathds 1^2}) = 0
\end{align}
where $\mathds 1 = e^{\frac{e_I}{\hbar c}\int^x_\mathcal{C} A^{\mu'} dx_{\mu'}}$ is the parallel-transported scale from space-time point $x_0 \to x$ along the particle trajectory $\mathcal C$, defined by the guidance equation
\begin{align}
    v^i \equiv \frac{j^i}{j^0} = \frac{\overline \psi  \gamma^i \psi}{\overline \psi \gamma^0  \psi} \label{dguy}
\end{align}
Note that \eqref{dguy} implies that the form of the guidance equation is unchanged by the introduction of $e_I$, as in section \ref{nonrel}. Also note that the definition of the scale factor $\mathds 1$ is the same for the Schrödinger, Pauli and Dirac equations. \\

Let us check whether our results are local scale invariant. Under a general gauge transformation $A^\mu \to A^\mu -\partial^\mu \lambda$, the weight factor transforms as $\mathds 1^2 \to \mathds 1'^2 = \mathds 1^2 e^{-\frac{2e_I}{\hbar c}\lambda }$ (see section \ref{schrcon}), the metric transforms as $g_{\mu \nu}\to g'_{\mu \nu} = g_{\mu \nu}e^{-\frac{2e_I\lambda}{\hbar c}}$, the metric determinant transforms as $\sqrt{-g}\to \sqrt{-g'} = \sqrt{-g} e^{-\frac{4e_I\lambda}{\hbar c}}$, the gamma matrices transform as $\gamma^\mu \to \gamma'^\mu = \gamma^\mu e^{\frac{e_I\lambda}{\hbar c}}$, and the quantum state transforms as $\psi \to \psi' = \psi e^{i\frac{e\lambda}{\hbar c} + \frac{1}{2}\frac{e_I\lambda}{\hbar c}}$(see appendix \ref{diraa}). Therefore, $\sqrt{-g} \textbf{ }\overline \psi  \gamma^\mu \psi \to \sqrt{-g'} \textbf{ }\overline {\psi'}  \gamma'^\mu \psi' = \sqrt{-g'} \textbf{ }\overline \psi  \gamma^\mu \psi e^{-\frac{2e_I}{\hbar c}\lambda }$, which implies that the conserved 4-current $\sqrt{-g} \textbf{ }\overline \psi  \gamma^\mu \psi/\mathds 1^2$ is gauge invariant, that is, both local scale and local phase invariant. The guidance equation (\ref{dguy}) is trivially invariant under gauge transformations. \\

Lastly, let us check the local scale invariance of the Dirac equation itself. We know that the Dirac spinor has a Weyl weight of $-3/2$ in the case of no gauge coupling and zero mass \cite{daviescurvy}. This implies that the spinor connection $\Gamma_\mu$ transforms as $\Gamma_\mu \to \Gamma'_\mu = \Gamma_\mu  - \hat I_4(3e_I)\partial_\mu \lambda/(2\hbar c) $ under a Weyl rescaling of the metric. Therefore, under a general gauge transformation, the Dirac equation \eqref{diro} transforms as 
\begin{align}
      \gamma^\mu  e^{\frac{e_I\lambda}{\hbar c}}\bigg (i\hbar \big (\hat I_4\partial_\mu + \Gamma_\mu - \hat I_4 \frac{3e_I}{2\hbar c}\partial_\mu \lambda) -  \hat I_4\frac{e_C}{c}\big (A_\mu - \partial_\mu \lambda\big )  \bigg )\psi e^{i\frac{e\lambda}{\hbar c} + \frac{1}{2}\frac{e_I\lambda}{\hbar c}} = mc \psi e^{i\frac{e\lambda}{\hbar c} + \frac{1}{2}\frac{e_I\lambda}{\hbar c}} \nonumber \\
      \Rightarrow \gamma^\mu  e^{\frac{e_I\lambda}{\hbar c}}\bigg (i\hbar \nabla_\mu -  \hat I_4\frac{e_C}{c}A_\mu + \hat I_4 \partial_\mu \lambda \big (\frac{e}{c} -  \frac{ie_I}{2 c}\big )\bigg )\psi e^{i\frac{e\lambda}{\hbar c} + \frac{1}{2}\frac{e_I\lambda}{\hbar c}} = mc \psi e^{i\frac{e\lambda}{\hbar c} + \frac{1}{2}\frac{e_I\lambda}{\hbar c}} \label{lsi}
\end{align}
Equation \eqref{lsi} implies that the Dirac equation is always local phase invariant, but is local scale invariant only in the massless case $m = 0$. Further, \eqref{lsi} clarifies that the transformations of both $\Gamma_\mu$ and $A_\mu$ jointly lead to the Weyl weight $-1/2$ for the quantum state. Note that the current density and the guidance equation, which together encode all physical predictions, are always local scale invariant.

\section{Axion-electrodynamics}\label{axion}
In sections \ref{nonrel}, \ref{multiple}, \ref{pauli} and \ref{dirac}, we have considered quantized particles interacting with an external electromagnetic field. This may be considered limited as, first, the number of particles is fixed in particle quantization and, second, the electromagnetic field is treated classically. In this section, we remove these limitations by extending our approach to axion electrodynamics, where a quantized axion field interacts with a quantized electromagnetic field. We remove infinities throughout our calculations.\\

The Lagrangian density for the system is given by 
\begin{align}
    \mathcal{L} = -\frac{1}{4}F^{\mu \nu} F_{\mu \nu} + \frac{1}{2}\big( \partial^\mu a \partial_\mu a - \frac{m^2 c^2}{\hbar^2}a^2 \big) + \frac{1}{4} g_C a F_{\mu \nu}\tilde{F}^{\mu \nu} \label{lag}
\end{align}
where $g_C= g + i g_I$ is a complex coupling parameter between the real axion field $a(x)$ and the electromagnetic gauge field $A^\mu (x)$, $F^{\mu \nu} = \partial^\mu A^\nu - \partial^\nu A^\mu$ and $\tilde{F}^{\mu \nu} = \frac{1}{2} \epsilon^{\mu \nu \sigma \lambda}F_{\sigma \lambda}$. \\

We quantize the system in Weyl gauge. The functional Schrödinger equation can be shown to be (see appendix \ref{axio1})
\begin{align}
&\int_\mathcal M \frac{1}{2} \bigg \{\big [ 
(-i\hbar c \frac{\delta }{\delta a}  + g_C\vec A \cdot \vec{B}) \cdot (-i\hbar c \frac{\delta }{\delta a} + g_C \vec A \cdot\vec{B})+ (\vec{\nabla} a)^2 + \frac{m^2c^2}{\hbar^2} a^2\big ] \nonumber \\
& + \big [ (-i \hbar c\frac{\delta }{\delta \vec A} + g_Ca \vec B) \cdot (-i \hbar c\frac{\delta }{\delta \vec A} +g_Ca \vec B) +\vec{B}^2\big ] \bigg \} \Psi[a, \vec A, t] = i \hbar\frac{\partial \Psi[a, \vec A, t]}{\partial t} \label{axmain}
\end{align}
where $\mathcal M$ labels the spatial manifold. The continuity equation can be shown to be (see appendix \ref{axio2})
\begin{align}
    \frac{\partial}{\partial t} \bigg (\frac{R^2}{\mathds 1^2}\bigg) + \int_\mathcal M \frac{\delta }{\delta a}\bigg( \frac{R^2}{\mathds 1^2}(c\frac{\delta S}{\delta a} + g\vec A \cdot \vec B)c\bigg) + \int_\mathcal M\frac{\delta }{\delta \vec A} \cdot \bigg(\frac{R^2}{\mathds 1^2}(c\frac{\delta S}{\delta \vec A} + ga\vec B)c\bigg) = 0
\end{align}
where $\mathds 1 = e^{\frac{g_I}{\hbar c}\int_\mathcal{C}^{(a, \vec A)} (\int_\mathcal M \vec A' \cdot \vec B' \delta a') + (\int_\mathcal M a' \vec B' \cdot \delta \vec A')}$ is the parallel-transported scale from $[a_0, \vec A_0, t_0] \to [a, \vec A, t]$ along the system trajectory $\mathcal{C} = \{\big (a'(\lambda), \vec A'(\lambda), t'(\lambda)\big)| \lambda \in [0,1]\}$, defined by the guidance equations
\begin{align}
    \frac{\partial a}{\partial ct} &= c\frac{\delta S}{\delta a} + g\vec A \cdot \vec B \label{guy1}\\
    \frac{\partial \vec A}{\partial ct} &= c\frac{\delta S}{\delta \vec A} + ga\vec B \label{guy2}
\end{align}
which determine the evolution of the axion and the gauge fields. Note that $\mathcal C$ is defined in the configuration space of field configurations. The Hamilton-Jacobi equation is
\begin{align}
    \frac{\partial S}{\partial t} + &\frac{1}{2} \int_\mathcal M \bigg \{ \bigg[(c\frac{\delta S}{\delta a} + g\vec A \cdot \vec B )^2 + (\vec \nabla a)^2 + \frac{m^2c^2}{\hbar^2}a^2 \bigg] + \frac{1}{2} \bigg [(c\frac{\delta S}{\delta \vec A} + g a\vec B)^2 + \vec B^2\bigg] \bigg\} \nonumber \\
    & -\frac{\hbar^2c^2}{2}\int_\mathcal M \bigg \{\frac{1}{R}\frac{D^2 R}{D a^2} + \frac{1}{R} \frac{D^2 R}{D \vec{A}^2} \bigg\} = 0 \label{hjax'}
\end{align}
where $D/Da$, $D/D\vec A$ are the Weyl covariant field derivatives (see appendix \ref{axio2}). It can be verified that our formulation is gauge invariant (see appendix \ref{axio3}).

\section{Pilot-wave equilibrium and uniqueness of trajectories}\label{prob}
In PWT, the probability density for an ensemble of configurations $\rho$ is conceptually distinct from the current density obtained from the Schrödinger equation \cite{bohm1, bohm2, bohm54, bohmbook2}. The continuity equation for an arbitrary $\rho$ is given by
\begin{align}
     \frac{\partial \rho }{\partial t} + \vec \nabla \cdot (\rho \vec v) = 0 \label{konti}
\end{align}
where the velocity field $\vec v$ is given by the guidance equation determined from the quantum state. Note that $\rho$ must be local scale invariant in our theory. Therefore, it may also be expressed as the local ratio $\rho \equiv \tilde \rho/\mathds 1^2$, where $\tilde \rho$ is gauge dependent and scales as $\tilde \rho \to \tilde \rho e^{-\frac{2e_I \lambda}{\hbar c}}$ upon a gauge transformation $A^\mu \to A^\mu - \partial^\mu \lambda $. Clearly, the gauge invariant $\rho$ is the physical probability density.\\

\subsection{Pilot wave equilibrium} \label{pweq}
For normalizable quantum states evolving via a Hermitian hamiltonian, the quantum predictions are reproduced if $\rho = R^2$, which is known as the quantum equilibrium condition. Bohm argued that the chaotic nature of the deterministic pilot-wave trajectories leads to relaxation from an arbitrary $\rho \to R^2$ at a coarse-grained level \cite{bohmbook2relax}. This idea was further developed by Valentini \cite{valentinI, royalvale}, whereas a different approach using the notion of typicality was taken by D{\"u}rr \textit{et al.} \cite{daro}. \\

In our scenario, we cannot directly identify the equilibrium condition by reference to orthodox quantum theory, as the hamiltonian is non Hermitian. The notion of quantum equilibrium is, however, limited as it imposes a condition borrowed from orthodox quantum theory into PWT. The notion internal to PWT is pilot-wave equilibrium \cite{sen22}, which applies to a wider range of conditions, such as non-normalizable states and non-Hermitian hamiltonians. This notion of equilibrium is applicable regardless of whether the predictions from orthodox quantum theory, if well-defined, are reproduced. Pilot-wave equilibrium reduces to quantum equilibrium when appropriate conditions are satisfied. The notion of pilot-wave equilibrium has recently been applied to the case of the non-normalizable Kodama state in non-perturbative quantum gravity \cite{sentarity}.\\

It is straightforward to apply the notion of pilot-wave equilibrium to our scenario. We note that the continuity equation
\begin{align}
    \frac{\partial}{\partial t}(\frac{R^2}{ \mathds 1^2}) + \vec{\nabla}\cdot (\frac{R^2}{ \mathds 1^2} \vec v) = 0
\end{align}
and equation (\ref{konti}) imply that
\begin{align}
    \frac{d}{dt} \bigg(\frac{\rho}{(R^2/\mathds 1^2 )}\bigg) = 0 \label{traj}
\end{align}
where $d/dt \equiv \partial /\partial t + \vec v \cdot \vec \nabla $ is the total time derivative. Equation (\ref{traj}) implies that the ratio $\rho / (R^2/\mathds 1^2)$ remains constant along the trajectories. We can then define pilot-wave equilibrium for our scenario to be satisfied if this ratio is unity, in which case the equilibrium probability density is given by 
\begin{align}
\rho_{eq} \equiv \frac{R^2}{\mathds 1^2}
\end{align}
at all times.\\

We define the $H$-function as the relative entropy between an arbitrary $\rho$ and the pilot-wave equilibrium density:
\begin{align}
H_{pw} \equiv \int_{\mathds C} \rho \ln \bigg (\frac{\rho}{R^2/ \mathds 1^2[\mathcal C]} \bigg)  \label{h}
\end{align}
where $\mathds C$ is the configuration space. It is straightforward to verify that the $H$-function (\ref{h}) satisfies all criteria for pilot-wave equilibrium outlined in \cite{sen22}. Therefore, an arbitrary density $\rho$ will tend to relax to $R^2/ \mathds 1^2$ (equivalently, $\tilde \rho$ will tend to relax to $R^2$) at a coarse-grained level, subject to the usual thermodynamic assumptions about initial conditions \cite{davies77}.

\subsection{Uniqueness of trajectories} \label{uniq}
As PWT provides a trajectory-based description of quantum phenomena, it is in general important to determine the uniqueness of the trajectories. For the spinless systems considered, the quantum Hamilton-Jacobi equation unambiguously fixes the conjugate momenta involved, and thereby the guidance equations. For the spin-1/2 systems considered, the relativistic current is uniquely fixed by the Dirac equation \cite{holland99}. The non-relativistic limit of the Dirac current then uniquely fixes the guidance equation for the Pauli equation \cite{bohmbook2, holland99}. Therefore, the velocity field is unambiguously fixed by the internal structure of PWT in both the Hermitian and non-Hermitian cases for the systems considered in this paper.\\
 
However, in the Hermitian case, alternative velocity fields generate the same experimental predictions in quantum equilibrium if they lead to conservation of the current density $R^2$. These alternative velocity fields can be obtained by adding a divergence-less term to the quantum current. We show below that, in the non-Hermitian case, this is no longer possible. That is, there is a one-to-one relationship between the velocity field and the equilibrium probability density -- changing one unambiguously changes the other.\\
 
Consider the pair of current density and velocity field $\big ( R^2, (\vec \nabla S - e \vec A)/m \big )$ in equation (\ref{conya2}). Consider changing the velocity field from $(\vec \nabla S - e \vec A)/m \to (\vec \nabla S - e \vec A)/m + \vec v'$, where $\vec v'$ is a gauge-invariant addition to the velocity field. For the added current $R^2\vec v'$ to have zero Weyl divergence, we must have
\begin{align}
    \vec D \cdot (R^2\vec v') = (\vec \nabla + \frac{2e_I}{\hbar c}\vec A) \cdot (R^2\vec v') = 0 \label{unix}
\end{align}
It is straightforward to show that the continuity equation with respect to ordinary derivatives, analogous to equation (\ref{finale}), is then modified to 
\begin{align}
     \frac{\partial}{\partial t}(\frac{R^2}{\mathds 1^2[\mathcal C']}) + \vec{\nabla}\cdot \bigg(\frac{R^2}{\mathds 1^2[\mathcal C']} \big (\frac{\vec{\nabla}S - e \vec{A}/c}{m} + \vec v' \big)\bigg) = 0
\end{align}
where the parallel-transported scale to measure $R$ has changed from $\mathds 1[\mathcal C] \to \mathds 1[\mathcal C']$, where $\mathcal C'$ ($\mathcal C$) is the trajectory determined by the velocity field $(\vec \nabla S - e \vec A)/m + \vec v'$ $\big ((\vec \nabla S - e \vec A)/m \big )$ and $\mathds 1[\mathcal C'] = e^{\frac{e_I}{\hbar c}\int^x_\mathcal{C'} A^{\mu'} dx_{\mu'}}$ ($\mathds 1[\mathcal C] = e^{\frac{e_I}{\hbar c}\int^x_\mathcal{C} A^{\mu'} dx_{\mu'}}$). The current density is changed from $R^2/\mathds 1^2[\mathcal C] \to R^2/\mathds 1^2[\mathcal C']$. Note that the form of the current density remains invariant, as in the Hermitian case.\\

The experimental predictions are then changed in pilot-wave equilibrium, as the equilibrium probability density $\rho_{eq}$ changes from $R^2/\mathds 1^2[\mathcal C] \to R^2/\mathds 1^2[\mathcal C']$. We conclude that the internal structure of PWT unambiguously determines the trajectories, and further that it is not possible, in the non-Hermitian case, to modify the trajectories without changing the equilibrium probability density. \\

\section{Discussion} \label{disc}
We have shown that the internal structure of PWT contains a hitherto unnoticed fundamental symmetry: local scale invariance. This renders the amplitude and phase of the quantum state symmetrical in the sense that both local scale invariance and local phase invariance hold true in the theory. Our results are general, applying from non-relativistic quantum mechanics through quantum field theory. The price for this new symmetry is non-Hermiticity for closed systems, which is prohibited in orthodox quantum theory but naturally realized in PWT. We have shown that the equilibrium probability density in non-Hermitian PWT is trajectory dependent, rendering the theory experimentally distinguishable from orthodox quantum theory and other quantum formulations in principle. We explore the experimental predictions of our theory in a subsequent paper \cite{sen26b}. \\

A key feature enabling local scale invariance in non-Hermitian PWT is that pilot-wave dynamics is formulated in configuration space. Trajectories governed by the guidance equation cannot cross in configuration space, so that each point in the support of the quantum state lies on a single pilot-wave trajectory. Therefore, the scale factor $\mathds 1[\mathcal{C}]$ is single-valued. In phase space based dynamics, by contrast, multiple trajectories with different momenta can pass through the same configuration point, rendering the scale factor, and thereby the current density, multi-valued. This observation bears on a common criticism of PWT, namely that it privileges the configuration-space representation \cite{pauli53, heisenbop, myrvohm}. Our work suggests that the configuration-space formulation is not an arbitrary choice but a key structural component that makes local scale invariance possible. It may be said that non-Hermitian PWT trades the representation symmetry in orthodox quantum theory for local scale symmetry. \\

Although we have considered only the electromagnetic gauge field, our work can be straightforwardly generalized to other gauge fields in the standard model by non-Hermitian extensions of the gauge covariant derivatives to accommodate local scale invariance. This suggests that the current standard model is an approximation to a non-Hermitian, locally scale invariant theory. Such local scale invariant extensions of the standard model have been proposed in the literature in various contexts \cite{rajpoop, turok14, scholz18}. The primary novelty of our approach lies in providing a general, fully quantized realization, which is challenging to achieve using orthodox quantum theory (see, for example, refs. \cite{codello, pagal}).\\

Our work challenges and nuances the received view that mass terms break local scale invariance. It shows that although mass terms may break the local scale invariance of formal quantities, physical quantities defined as local ratios -- such as the current density, the guidance equation, and the quantum potential -- always remain local scale invariant. This is particularly clearly brought out by the case of the Dirac equation in curved spacetime (see section \ref{dirac}). The Dirac equation itself is local scale invariant only in the massless case, but all the physical predictions, determined by the current density and the guidance equation, remain local scale invariant even in the massive case. In the subsequent paper \cite{sen26b}, we have shown that the role of mass is to set the magnitude of non-integrable scale effects. \\

Our work suggests that it is promising to quantize the Weyl action \cite{weyl22} in pilot-wave approaches to quantum gravity. As of yet, quantum gravity works using PWT are solely focused on the Einstein-Hilbert action (see, for example, refs. \cite{appraisal, shtanov96, shajai3, pintu13, pintu19, sentarity}) which is not local scale invariant. Given the unique trajectory-dependent definition of the inner product in non-Hermitian PWT, we can expect interesting answers to issues in quantizing the Weyl action in the orthodox approach, such as ghost (negative norm) states \cite{ghost1, ghost2}. \\

The problem of non-uniqueness of trajectories in Hermitian PWT \cite{98holland, uniquecollapse, finkel99}  vanishes in the non-Hermitian case (see section \ref{uniq}). The guidance equation is uniquely determined by the Hamilton-Jacobi equation (relativistic considerations) in the spinless (spin) case. Adding a divergence-less term to the current modifies the guidance equation, which in turn modifies the equilibrium probability density $|\psi|^2/\mathds 1^2[\mathcal C]$ due to its dependence on the trajectory $\mathcal C$. Lastly, we have shown that the current for the Pauli equation in the non-Hermitian case contains the term $(\vec D \times \vec s)/m$, which reduces, in the Hermitian case, to the familiar spin-current term $(\vec \nabla \times \vec s)/m$ that can be derived from the non-relativistic limit of the Dirac equation \cite{holland99}. Our derivation of this term provides a new proof based directly on the Pauli equation. \\

Our work suggests the possibility of defining a time operator in PWT. In previous works we have shown that normalizability is not inherent in the conceptual structure of PWT \cite{sen22, sentarity}. With both normalizability and Hermiticity removed, Pauli's objection to the existence of a time operator in orthodox quantum theory \cite{pauli80} is no longer applicable to PWT. Such a non-Hermitian time operator may lead to trajectory-dependent predictions, similar to the equilibrium probability density $|\psi|^2/\mathds 1^2$. We leave to future work to determine the relation, if any, between such a time operator and the trajectory-dependent arrival times explored in the pilot-wave literature \cite{leave, mousa, das19, das25}.\\

We make a few remarks on the philosophical implications of our work. PWT has often been presented within a scientific realist setting, whereas Weyl presented his gauge theory in the context of Husserl's transcendental idealism \cite{ryckmanbook}. This suggests that a confluence of ideas from both scientific realism and transcendental idealism may be best suited to understand our work from a philosophical perspective. There may appear to be a conflict between Weyl's idea of infinitesimal, or purely local, geometry and the nonlocality of pilot-wave dynamics. However, we have implemented the infinitesimal idea in configuration space, where pilot-wave dynamics is local. This is related to Bohm's idea of implicate order \cite{bohmbook2, bohmplicate, bohman}, where we may consider the space-time dynamics of particles to be an unfolding of particle configuration space dynamics, which itself may be considered to be an unfolding of field configuration space dynamics, potentially ad infinitum. Our work suggests that Weyl's idea of ``pure infinitesimal geometry'' \cite{weyl18} is applicable at each level of this order.\\

It is tempting to view our theory as a generalization of PWT to incorporate local scale invariance and non-Hermiticity. However, treating PWT as theory in its own right suggests that we are merely uncovering its internal conceptual structure. This structure was illuminated to a great extent by Bohm, but in an era where orthodox quantum theory was the established dogma. This led to several features being uncritically imported from orthodox quantum theory into PWT. The theory we currently know as the standard, Hermitian PWT is the version wherein its local scale invariance symmetry is omitted by hand -- the result of importing the reinterpretation, made in the 1920s, of Weyl's gauge idea as local phase invariance in orthodox quantum theory \cite{fuck, london27, vizgin, gawn}. Consequently, the trajectories did not affect the equilibrium probability density, leading to the criticism that they constitute ``ideological superstructures'' \cite[pp. 145-146]{heisenbop}. From its internal perspective, the theory is non-Hermitian, local scale invariant as well as local phase invariant, and the trajectories are not hidden but enter explicitly into the definition of the probability rule.

\section*{Author Contributions}
Indrajit Sen: conceptualization (lead); methodology (lead); formal analysis (lead); supervision (equal); validation (equal); writing - original draft (lead); writing - review \& editing (equal). Matt Leifer: conceptualization (supporting); methodology (supporting); supervision (equal); validation (equal); writing - review \& editing (equal).

\acknowledgments
IS is thankful to Krzysztof Sienicki for helpful comments on an earlier version of the manuscript, and to Sayani Ghosh for helpful discussions. ML was supported by Grant 63209 from the John Templeton Foundation. The opinions expressed in this publication are those of the authors and do not necessarily reflect the views of the John Templeton Foundation.

\bibliographystyle{bhak}
\bibliography{bib}

\appendix

\section{Multiple spinless non-relativistic particles}\label{multipla}
Using the polar decomposition ansatz $\psi(\vec{X},t) = R(\vec{X},t) e^{i S(\vec{X},t)/\hbar}$, the imaginary part of the Schrödinger equation (\ref{schrn}) can be shown to be
\begin{align}
     \frac{\partial R^2}{\partial t} + \sum_{j= 1}^N\vec{\nabla}_j\cdot (R^2 \frac{\vec{\nabla}_jS - e_j \vec{A}_j/c}{m_j}) = 2R^2\sum_{j= 1}^N\bigg(-\frac{e_{Ij}}{\hbar c} \vec{A}_j\cdot \frac{(\vec{\nabla}_jS - e_j \vec{A}_j/c)}{m_j}  + \frac{e_{Ij}}{\hbar} \phi_j\bigg)
\end{align}
which can be rewritten as the continuity equation
\begin{align}
    D_0 (cR^2) + \sum_{j=1}^N\vec D_j \cdot (R^2 \frac{\vec \nabla_j S - e_j \vec A_j/c}{m_j}) = 0 \label{conyan}
\end{align}
where $X^0 \equiv ct$ and we have replaced the ordinary derivatives by the Weyl covariant derivatives (in configuration space)
\begin{align}
    &\partial_0 \to D_0 \equiv (\partial_0 - \sum_{j=1}^N \frac{\omega e_{Ij}}{\hbar c} \phi_j) \label{coven1} \\
    &\vec \nabla_j \to \vec D_j \equiv (\vec \nabla_j + \frac{\omega e_{Ij}}{\hbar c} \vec{A}_j) \label{coven2}
\end{align}
acting on $R^2$, which has Weyl weight $\omega = 2$, determined by the transformation $R^2 \to R'^2 = R^2 \prod_{j= 1}^N e^{-\frac{2e_I \lambda_j}{\hbar c}}$. Analogous to the case for a single particle, let us suppose that the density conserved with respect to the ordinary derivatives is $R^2/\Omega$, where $\Omega$ is a weight factor to be determined. Equation (\ref{conyan}) implies that
\begin{align}
    \frac{\partial}{\partial t}\frac{R^2}{\Omega} + &\sum_{j=1}^N\vec \nabla_j \cdot (\frac{R^2}{\Omega} \frac{\vec \nabla_j S - e_j \vec A_j/c}{m_j}) = \nonumber \\
    &R^2 \sum_{j=1}^N \bigg(\big (-\frac{\vec{\nabla}_j\Omega}{\Omega^2}-2\frac{e_{Ij}}{\Omega \hbar c} \vec{A}_j)\cdot \frac{(\vec{\nabla}_jS - e_j \vec{A}_j/c)}{m_j}  + \big (-\frac{1}{\Omega^2 N}\frac{\partial \Omega }{\partial t} + \frac{2e_{Ij}}{\hbar\Omega} \phi_j \big )   \bigg) 
\end{align}

Clearly, $R^2/\Omega$ will be conserved if 
\begin{align}
    \sum_{j=1}^N \bigg(\big (-\frac{\vec{\nabla}_j\Omega}{\Omega^2}-2\frac{e_{Ij}}{\Omega \hbar c} \vec{A}_j)\cdot \frac{(\vec{\nabla}_jS - e_j \vec{A}_j/c)}{m_j}  + \big (-\frac{1}{\Omega^2 N}\frac{\partial \Omega }{\partial t} + \frac{2e_{Ij}}{\hbar\Omega} \phi_j \big )   \bigg)  = 0 \label{constrn}
\end{align}
which can be rewritten as 
\begin{align}
     \sum_{j=1}^N \vec v_j \cdot \vec D_j \Omega + v^0 D_0 \Omega = 0 \label{ncons}
\end{align}
where $v^0 \equiv dX^0/dt = c$ and the Weyl covariant derivatives are defined by (\ref{coven1}), (\ref{coven2}). Equation (\ref{ncons}) has the path-dependent solution
\begin{align}
 \Omega = \Pi_{j=1}^N e^{\frac{2e_{Ij}}{\hbar c}\int^{(ct, \vec{X})}_\mathcal{C} A_{j\mu'} dx^{\mu'}_j} = \mathds 1^2[\mathcal C] \label{nolulu}
\end{align}
where $\mathcal{C} = \{ (t'(\lambda), \vec X'(\lambda))| \lambda \in [0,1] \}$ is the configuration space trajectory of the $N$-particle system and $e^{\frac{2e_{Ij}}{\hbar c}\int^{(ct, \vec{X})}_\mathcal{C} A_{j\mu'} dx^{\mu'}_j}$ is a line integral over the co-ordinates $(ct, \vec x_j)$ along $\mathcal C$. Analogous to section \ref{nonrel} in the main text, we have set $\Omega_0(\vec X_0, t_0) \equiv  1$ in the gauge which satisfies $R(\vec X_0, t_0) = \tilde R(\vec X_0, t_0)$, where $(\vec X_0, t_0)$ is the initial point of the curve $\mathcal C$ and $\tilde R(\vec X_0, t_0)$ is the initial gauge-invariant amplitude when $e_I = 0$.

\section{Spin-1/2 non-relativistic particle} \label{paula}
The 2 components of the Pauli equation (\ref{paul}) in the main text are
\begin{align}
    \big [\frac{(-i\hbar\vec{\nabla} -e_C\vec{A}/c) \cdot (-i\hbar\vec{\nabla} -e_C\vec{A}/c)}{2m} + e_C\phi\big] \psi_+(\vec{x},t)  - \frac{\hbar e_C}{2mc}\langle+|\vec B \cdot \hat{\vec \sigma}|\psi(\vec x, t) \rangle = i \hbar \frac{\partial \psi_+(\vec{x},t)}{\partial t} \label{paul1} \\
\big [\frac{(-i\hbar\vec{\nabla} -e_C\vec{A}/c) \cdot (-i\hbar\vec{\nabla} -e_C\vec{A}/c)}{2m} + e_C\phi\big] \psi_-(\vec{x},t)  - \frac{\hbar e_C}{2mc}\langle -|\vec B \cdot \hat{\vec \sigma}|\psi(\vec x, t) \rangle= i \hbar \frac{\partial \psi_-(\vec{x},t)}{\partial t} \label{paul2}
\end{align}

Using equations (\ref{paul1}), (\ref{paul2}) and their conjugates, we can obtain
\begin{align}
     \frac{\partial (R_+^2 + R_-^2)}{\partial t} + \vec{\nabla}\cdot (R_+^2 \frac{\vec{\nabla}S_+ - e \vec{A}/c}{m}+R_-^2 \frac{\vec{\nabla}S_- - e \vec{A}/c}{m}) =  \nonumber \\
     2R_+^2\bigg(-\frac{e_I}{\hbar c} \vec{A}\cdot \frac{(\vec{\nabla}S_+ - e \vec{A}/c)}{m}  + \frac{e_I}{\hbar} \phi \bigg) + 2R_-^2\bigg(-\frac{e_I}{\hbar c} \vec{A}\cdot \frac{(\vec{\nabla}S_- - e \vec{A}/c)}{m}  + \frac{e_I}{\hbar} \phi \bigg) - \frac{2e_I}{\hbar mc}  \vec B \cdot \vec s \label{conpa}
\end{align}
where 
\begin{align}
\vec s = \frac{\hbar}{2} \langle \psi(\vec x, t)|\hat{\vec \sigma}|\psi(\vec x, t) \rangle
\end{align}
and $ \psi_+(\vec x, t) = R_+(\vec x, t) e^{iS_+(\vec x , t)/\hbar}$, $ \psi_-(\vec x, t) = R_-(\vec x, t) e^{iS_-(\vec x , t)/\hbar}$. We observe that we can rewrite the last term in (\ref{conpa}) using 
\begin{align}
      \vec B \cdot \vec s = \vec \nabla \cdot (\vec A \times \vec s) + \vec A \cdot ( \vec \nabla \times \vec s) \label{thru}
\end{align}
where $\vec B = \vec \nabla \times \vec A$. Clearly, the term $\vec \nabla \cdot (\vec A \times \vec s)$ can be absorbed into the current to give 
\begin{align}
     \frac{\partial (R_+^2 + R_-^2)}{\partial t} + \vec{\nabla}\cdot (R_+^2 \frac{\vec{\nabla}S_+ - e \vec{A}/c}{m}+R_-^2 \frac{\vec{\nabla}S_- - e \vec{A}/c}{m} + \frac{2e_I}{\hbar mc}\vec A \times \vec s) =  \nonumber \\
     2R_+^2\bigg(-\frac{e_I}{\hbar c} \vec{A}\cdot \frac{(\vec{\nabla}S_+ - e \vec{A}/c)}{m}  + \frac{e_I}{\hbar} \phi \bigg) + 2R_-^2\bigg(-\frac{e_I}{\hbar c} \vec{A}\cdot \frac{(\vec{\nabla}S_- - e \vec{A}/c)}{m}  + \frac{e_I}{\hbar} \phi \bigg) - \frac{2e_I}{\hbar mc}  \vec A \cdot ( \vec \nabla \times \vec s) \label{conpao}
\end{align}
Equation (\ref{conpao}) implies that $\langle \psi(\vec x, t)|\psi(\vec x, t)\rangle \equiv R_+^2 + R_-^2$ is not a locally conserved density. Let us consider the alternate density $(R_+^2 + R_-^2)/\Omega$. Equation (\ref{conpao}) then implies that
\begin{align}
    &\frac{\partial}{\partial t}(\frac{R_+^2 + R_-^2}{\Omega}) + \vec{\nabla}\cdot \bigg (\frac{1}{\Omega} (R_+^2 \frac{\vec{\nabla}S_+ - e \vec{A}/c}{m}+R_-^2 \frac{\vec{\nabla}S_- - e \vec{A}/c}{m} + \frac{2e_I}{\hbar mc}\vec A \times \vec s)\bigg ) = (R_+^2 + R_-^2) \bigg ( -\frac{1}{\Omega^2}\frac{\partial \Omega }{\partial t} + \frac{2e_I}{\Omega \hbar} \phi \bigg ) \nonumber \\
    &+R_+^2 \bigg ((-\frac{\vec \nabla \Omega}{\Omega^2}-\frac{2 e_I}{\Omega \hbar c} \vec{A})\cdot \frac{(\vec{\nabla}S_+ - e \vec{A}/c)}{m}  \bigg) +R_-^2\bigg((-\frac{\vec \nabla \Omega}{\Omega^2}-\frac{2 e_I}{\Omega \hbar c} \vec{A})\cdot \frac{(\vec{\nabla}S_- - e \vec{A}/c)}{m} \bigg)   \nonumber \\
    & +\bigg((-\frac{\vec \nabla \Omega}{\Omega^2}-\frac{2 e_I}{\Omega \hbar c} \vec{A})\cdot \frac{2e_I}{\hbar mc}(\vec A \times \vec s) \bigg) - \frac{2e_I}{\Omega \hbar mc}  \vec A \cdot ( \vec \nabla \times \vec s)  \label{conpa1}
\end{align}
where we have used $\vec A\cdot (\vec A \times \vec s) = 0$. Note that the term $\vec A \cdot ( \vec \nabla \times \vec s)$ on the right hand side of (\ref{conpa1}) is not eliminated by the introduction of $\Omega$ alone. This suggests that we modify the current in (\ref{conpao}) by a divergence-less additional term $(\vec \nabla \times \vec s)/m $ to eliminate $\vec A \cdot ( \vec \nabla \times \vec s)$. Equation (\ref{conpa1}) is then modified to 
\begin{align}
   &\frac{\partial}{\partial t}(\frac{R_+^2 + R_-^2}{\Omega}) + \vec{\nabla}\cdot \bigg (\frac{1}{\Omega} (R_+^2 \frac{\vec{\nabla}S_+ - e \vec{A}/c}{m}+R_-^2 \frac{\vec{\nabla}S_- - e \vec{A}/c}{m} +\frac{\vec \nabla \times \vec s}{m}+ \frac{2e_I}{\hbar mc}\vec A \times \vec s)\bigg ) = (R_+^2 + R_-^2) \bigg ( -\frac{1}{\Omega^2}\frac{\partial \Omega }{\partial t} + \frac{2e_I}{\Omega \hbar} \phi \bigg ) \nonumber \\
    &+R_+^2 \bigg ((-\frac{\vec \nabla \Omega}{\Omega^2}-\frac{2 e_I}{\Omega \hbar c} \vec{A})\cdot \frac{(\vec{\nabla}S_+ - e \vec{A}/c)}{m}  \bigg) +R_-^2\bigg((-\frac{\vec \nabla \Omega}{\Omega^2}-\frac{2 e_I}{\Omega \hbar c} \vec{A})\cdot \frac{(\vec{\nabla}S_- - e \vec{A}/c)}{m} \bigg)   \nonumber \\
    & +\bigg((-\frac{\vec \nabla \Omega}{\Omega^2}-\frac{2 e_I}{\Omega \hbar c} \vec{A})\cdot \frac{2e_I}{\hbar mc}(\vec A \times \vec s) \bigg) + \bigg ( (-\frac{\vec \nabla \Omega}{\Omega^2}-\frac{2 e_I}{\Omega \hbar c} \vec{A}) \cdot \frac{ \vec \nabla \times \vec s}{m} \bigg) \label{kunpau}
\end{align}
For $R^2/\Omega$ to be conserved, the right hand side of \eqref{kunpau} needs to vanish. This condition can be concisely expressed as
\begin{align}
    v^\mu D_\mu \Omega = 0 \label{gstring}
\end{align}
where the Weyl covariant derivatives are defined by equation (\ref{cove}) in the main text, $v^0 \equiv dx^0/dt = c $ and 
\begin{align}
    \vec v \equiv \frac{1}{m}\frac{R_+^2 (\vec{\nabla}S_+ - e \vec{A}/c)+R_-^2 (\vec{\nabla}S_- - e \vec{A}/c) +\vec D \times \vec s }{R_+^2 + R_-^2}
\end{align}
Clearly, the constraint \eqref{gstring} is identical in form to \eqref{gconstr} in the main text. It is straightforward to check that the solution to $\Omega$ is given by the path-dependent solution (\ref{solulu}) in the main text. 
\section{Spin-1/2 relativistic particle} \label{diraa}
The Dirac equation (\ref{diro}) and its adjoint are given by
\begin{align}
    \gamma^\mu (i\hbar \nabla_\mu - \frac{e+ie_I}{c} A_\mu)\psi = mc \psi \nonumber\\
    \overline \psi\gamma^\mu (-i\hbar \overleftarrow \nabla_\mu - \frac{e-ie_I}{c} A_\mu) = mc \overline \psi \label{dir}
\end{align}
where $ \overleftarrow \nabla_\mu \equiv (\overleftarrow\partial_\mu -\Gamma_\mu)$ is the spacetime covariant derivative acting to the left, $\overline \psi \equiv \psi^\dagger \gamma^{\hat 0}$ and $\gamma^{\hat 0}$ labels the flat-space gamma matrix in the local Lorentz frame. Using the pair of equations (\ref{dir}), it is straightforward to show that 
\begin{align}
    \nabla_\mu ( \overline \psi \gamma^\mu\psi) = \frac{2e_I}{\hbar c} A_\mu \overline \psi \gamma^\mu\psi \label{dcur}
\end{align}
where $\nabla_\mu ( \overline \psi \gamma^\mu\psi) = (\overline \psi \overleftarrow\nabla_\mu)  \gamma^\mu\psi + \overline \psi(\nabla_\mu \gamma^\mu) \psi + \overline \psi  (\gamma^\mu\nabla_\mu \psi) $. Equation \eqref{dcur} can be written as 
\begin{align}
     \partial_\mu (\sqrt{-g} \textbf{ }\overline \psi \gamma^\mu\psi) = \sqrt{-g} \textbf{ }\frac{2e_I}{\hbar c} A_\mu \overline \psi \gamma^\mu\psi \label{tharak}
\end{align}
where we have used $\nabla_\mu ( \overline \psi \gamma^\mu\psi)  = (1/\sqrt{-g} )\partial_\mu (\sqrt{-g} \textbf{ }\overline \psi \gamma^\mu\psi)$. Equation \eqref{tharak} suggests that the current density $\sqrt{-g} \textbf{ }\overline \psi \gamma^\mu\psi$ has a Weyl weight of 2, which we can use to fix the scale transformation of $\psi$. Under a general gauge transformation $A_\mu \to A_\mu -\partial_\mu \lambda$, the metric transforms as $g_{\mu \nu}\to g'_{\mu \nu} = g_{\mu \nu}e^{-\frac{2e_I\lambda}{\hbar c}}$, the metric determinant transforms as $\sqrt{-g}\to \sqrt{-g'} = \sqrt{-g} e^{-\frac{4e_I\lambda}{\hbar c}}$, the gamma matrices transform as $\gamma^\mu \to \gamma'^\mu = \gamma^\mu e^{\frac{e_I\lambda}{\hbar c}}$. Suppose the quantum state transforms as $\psi \to \psi' = \psi e^{i\frac{e\lambda}{\hbar c} - s\frac{e_I\lambda}{\hbar c}}$, where $s$ is a real number. Then, from the requirement that the Weyl weight of $\sqrt{-g}\textbf{ }\overline \psi \gamma^\mu\psi$ is 2, we get 
\begin{align}
   -4 -2s + 1 = -2 \nonumber \\
    \Rightarrow s = -\frac{1}{2}
\end{align}
Therefore, the quantum state transforms as $\psi \to \psi' = \psi e^{i\frac{e\lambda}{\hbar c} + \frac{1}{2}\frac{e_I\lambda}{\hbar c}}$. Analogous to section \ref{nonrel}, we can express equation (\ref{dcur}) as 
\begin{align}
    D_\mu (\sqrt{-g}\textbf{ } \overline \psi \gamma^\mu \psi) = 0 \label{dconya}
\end{align}
where $D_\mu$ is the Weyl covariant derivative, defined by \eqref{cove} in the main text, acting on $\sqrt{-g} \textbf{ }\overline \psi \gamma^\mu \psi$, which has Weyl weight $\omega = 2$. Equation (\ref{dconya}) implies that $\sqrt{-g} \textbf{ }\overline \psi \gamma^\mu \psi$ is the conserved density with respect to the Weyl covariant derivatives. Let us suppose that the density conserved with respect to the ordinary derivatives is $\sqrt{-g} \textbf{ }\overline \psi \gamma^\mu \psi /\Omega$, where $\Omega$ is a weight factor to be determined. Equation (\ref{dconya}) then implies that
\begin{align}
    \partial_\mu (\sqrt{-g} \textbf{ }\frac{\overline \psi \gamma^\mu\psi}{\Omega}) =  -\sqrt{-g} \textbf{ }\frac{\overline \psi \gamma^\mu \psi}{\Omega^2} D_\mu \Omega \label{dom1}
\end{align}
For the right hand side of (\ref{dom1}) to vanish, we must have 
\begin{align}
    v^\mu D_\mu \Omega = 0 \label{gstring1}
\end{align}
where $v^\mu \equiv \overline \psi \gamma^\mu \psi/(\overline \psi \gamma^0  \psi)$. Equation \eqref{gstring1}, which is identical in form to \eqref{gconstr} in the main text, has the path-dependent solution given by \eqref{solulu} in the main text.

\section{Axion Electrodynamics}
\subsection{Quantization} \label{axio1}
We work in the Weyl gauge $A^0 = 0$. The electric and magnetic fields are $\vec{E} = - \partial \vec{A}/\partial x^0$, $\vec{B}= \vec{\nabla} \times \vec{A}$ respectively. From the Lagrangian (\ref{lag}) in the main text, the canonical momentum conjugate to $a$ is $\pi_a = \partial \mathcal L/\partial \dot a =  \partial_0 a/c$, whereas the canonical momenta conjugate to $A^\mu$ are $\pi_{A^0} = \partial \mathcal L/\partial \dot A^0 = 0$, $\vec{\pi}_{\vec{A}} = \partial \mathcal L/\partial \dot{\vec{A}} = (-\vec{E} + g_Ca\vec{B})/c$. The Hamiltonian density is
\begin{align}
    \mathcal{H} = \frac{1}{2}\big [\pi_a^2c^2 + (\vec{\nabla} a)^2 + \frac{m^2 c^2}{\hbar^2} a^2 \big ] +\frac{1}{2}\big [ (\vec{\pi}_{\vec{A}}c - g_Ca\vec{B})^2 +\vec{B}^2\big ] \label{Ham1}
\end{align}
Let us apply a canonical transformation such that the canonical momenta of the gauge field and the axion field are symmetrically modified by the coupling. We apply the following generating functional of 2nd type
\begin{align}
    F_2[a, \vec A, \pi_a', \vec \pi_{\vec A}'] = \int_\mathcal M \textbf{ } (a\pi_a' + \vec{A}\cdot\vec{\pi}_{\vec{A}}' + \frac{g_Ca}{c} \vec{A}\cdot\vec B)
\end{align}
The transformed variables are given by the equations
\begin{align}
    \pi_a &= \frac{\delta F_2}{\delta a} = \pi_a' + \frac{g_C}{c}\vec{A}\cdot\vec{B} \\
\vec{\pi}_{\vec{A}} &= \frac{\delta F_2}{\delta \vec{A}} = \vec{\pi}_{\vec{A}}' + \frac{2g_C}{c}a \vec B\\
a' &= \frac{\delta F_2}{\delta \pi_a'} = a\\
\vec{A}' &= \frac{\delta F_2}{\delta \vec{\pi}'_{\vec{A}}} = \vec{A}
\end{align}
and the transformed Hamiltonian density is given by
\begin{align}
\mathcal{H}' = \mathcal{H} + \frac{\partial F_2}{\partial t} = \frac{1}{2}\big [(\pi'_a c + g_C\vec{A}\cdot\vec{B})^2 + (\vec{\nabla} a)^2 + \frac{m^2c^2}{\hbar^2} a^2 \big ] +\frac{1}{2}\big [ (\vec{\pi}_{\vec{A}}'c + g_Ca\vec B)^2 +\vec{B}^2\big ]\\
\end{align}

For quantization, we promote the fields and their canonical momenta to operators and impose the equal-time canonical commutation relations $[\hat{a}(\vec{x}, t), \hat{\pi}'_a(\vec{x}', t)] = i\hbar \delta(\vec{x}-\vec{x}')$, $[\hat{\vec{A}}(\vec x, t), \hat{\vec \pi'}_{\vec{A}} (\vec x', t)] = i \hbar \delta(\vec{x}-\vec{x}')$. Working in the $a, \vec A$ representation, we have $\hat{a} \to a$, $\hat{\pi}'_{a} \to -i\hbar \frac{\delta}{\delta a}$ and $\hat{\vec A} \to \vec A$, $\hat{\vec \pi}'_{\vec{A}} \to -i \hbar \frac{\delta }{\delta \vec A}$. \\

The functional Schrödinger equation is 
\begin{align}
& \int_\mathcal{M} \textbf{ } \hat{\mathcal{H}'} \Psi[a, \vec A] = i \hbar\frac{\partial \Psi[a, \vec A, t]}{\partial t} \\
    \Rightarrow &\int_\mathcal M \frac{1}{2} \bigg \{\big [ 
(-i\hbar c \frac{\delta }{\delta a}  + g_C\vec A \cdot \vec{B}) \cdot (-i\hbar c \frac{\delta }{\delta a} + g_C \vec A \cdot\vec{B})+ (\vec{\nabla} a)^2 + \frac{m^2c^2}{\hbar^2} a^2\big ] \nonumber \\
& + \big [ (-i \hbar c\frac{\delta }{\delta \vec A} + g_Ca \vec B) \cdot (-i \hbar c\frac{\delta }{\delta \vec A} +g_Ca \vec B) +\vec{B}^2\big ] \bigg \} \Psi[a, \vec A, t] = i \hbar\frac{\partial \Psi[a, \vec A, t]}{\partial t} \label{axschro}
\end{align}
where $\mathcal M$ labels the spatial manifold and $\int_\mathcal M \equiv \int_\mathcal M d^3\vec x$.

\subsection{Quantum Hamilton-Jacobi and Continuity equations} \label{axio2}
Using the polar decomposition $\Psi[a, \vec A, t] = R[a, \vec A, t] e^{iS[a, \vec A, t]/\hbar}$, the real part of the functional Schrödinger equation (\ref{axschro}) can be shown to be
\begin{align}
    \frac{\partial S}{\partial t} + &\frac{1}{2} \int_\mathcal M \bigg \{ \bigg[(c\frac{\delta S}{\delta a} + g\vec A \cdot \vec B )^2 + (\vec \nabla a)^2 + \frac{m^2c^2}{\hbar^2}a^2 \bigg] + \frac{1}{2} \bigg [(c\frac{\delta S}{\delta \vec A} + g a\vec B)^2 + \vec B^2\bigg] \bigg\} \nonumber \\
    & -\frac{\hbar^2}{2}\int_\mathcal M \bigg \{\frac{c^2}{R}(\frac{\delta^2 R}{\delta a^2} + \frac{\delta^2 R}{\delta \vec{A}^2}) + \frac{g_I^2}{\hbar^2} \big ( (\vec A \cdot \vec B)^2 + (a \vec B)^2 \big ) - 2\frac{g_I c}{\hbar R}(\vec A \cdot \vec B \frac{\delta R}{\delta a} + a \vec B \cdot \frac{\delta R}{\delta \vec A})\bigg\} = 0 \label{hjax}
    \end{align}
and its imaginary part can be shown to be
    \begin{align}
\frac{\partial R^2}{\partial t} + &\int_\mathcal M \frac{\delta }{\delta a}\bigg[R^2(c\frac{\delta S}{\delta a} + g\vec A \cdot \vec B)c\bigg] + \int_\mathcal M \frac{\delta }{\delta \vec A} \cdot \bigg[R^2(c\frac{\delta S}{\delta \vec A} + g a\vec B)c\bigg] = \nonumber \\ 
&2R^2\frac{g_I}{\hbar c} \int_\mathcal M  \bigg\{ \vec A \cdot \vec B \bigg[c\frac{\delta S}{\delta a} + g \vec A \cdot \vec B \bigg]c + a\vec B\cdot \bigg[c\frac{\delta S}{\delta \vec A} + g a \vec B \bigg]c \bigg\} \label{conjax}
\end{align}
Clearly, equation (\ref{hjax}) is the quantum Hamilton-Jacobi equation for the axion-electromagnetic system. It implies that the field momenta are given by $\pi'_a = \delta S/\delta a $ and $\vec \pi'_{\vec A} = \delta S/\delta \vec A$. The second line of equation (\ref{hjax}) contains the quantum potential, which has been modified by additional terms that depend on $g_I$. Equation (\ref{conjax}) is the quantum continuity equation, with terms on the right hand side that depend on $g_I$. These equations are analogous to the ones we obtained in sections \ref{nonrel} and \ref{dirac}. \\

Let us rewrite equation (\ref{conjax}) as
\begin{align}
    \frac{\partial R^2}{\partial t} + &\int_\mathcal M \textbf{ } \frac{D}{Da}\bigg[R^2(c\frac{\delta S}{\delta a} + g\vec A \cdot \vec B)c\bigg] + \int_\mathcal M \textbf{ }\frac{D}{D \vec A} \cdot \bigg[R^2(c\frac{\delta S}{\delta \vec A} + g a\vec B)c\bigg] = 0 \label{conyax}
\end{align}
where the ordinary field derivatives have been replaced by the Weyl covariant field derivatives
\begin{align}
    \frac{D}{Da} &\equiv \frac{\delta }{\delta a} -\frac{\omega g_I}{\hbar c}\vec A \cdot \vec B \label{fcove1}\\
    \frac{D}{D{\vec A}} &\equiv  \frac{\delta }{\delta \vec A} -\frac{\omega g_I}{\hbar c}a  \vec B \label{fcove2}
\end{align}
acting on $R^2$, which has Weyl weight $\omega =2$, determined by the transformation $R^2 \to R'^2 = R^2e^{\frac{2g_I}{\hbar c}\int_\mathcal{M} a\vec B \cdot \vec \nabla \lambda}$ upon an infinitesimal gauge transformation (see appendix \ref{axio3}). Clearly, the Weyl covariant deivatives depend on the imaginary component of the coupling parameter $g_I$. Equation (\ref{conyax}) implies that $R^2$ is locally conserved with respect to the Weyl covariant field derivatives.\\

Let us suppose the density conserved with respect to ordinary field derivatives is $R^2/\Omega $, where $\Omega$ is a weight factor to be determined. Equation (\ref{conjax}) then implies that
\begin{align}
    \frac{\partial}{\partial t}\frac{R^2}{\Omega} + &\int_\mathcal M \frac{\delta }{\delta a}\bigg[\frac{R^2}{\Omega}(c\frac{\delta S}{\delta a} + g\vec A \cdot \vec B)c\bigg] + \int_\mathcal M\frac{\delta }{\delta \vec A} \cdot \bigg[\frac{R^2}{\Omega}(c\frac{\delta S}{\delta \vec A} + ga\vec B)c\bigg] = \nonumber \\ 
&R^2\bigg\{ \int_\mathcal M \bigg[-\frac{1}{\Omega^2}\frac{\delta \Omega }{\delta a} + \frac{2 g_I}{\Omega\hbar c} \vec A \cdot \vec B\bigg] \bigg[c\frac{\delta S}{\delta a} + g\vec A \cdot \vec B \bigg]c + \int_\mathcal M \bigg[ -\frac{1}{\Omega^2}\frac{\delta \Omega}{\delta \vec A} + \frac{2 g_I}{\Omega\hbar c} a\vec B\bigg]\cdot \bigg[c\frac{\delta S}{\delta \vec A} + ga \vec B \bigg]c \bigg\} \label{n1}
\end{align}
For $R^2/\Omega$ to be conserved, the right hand side of equation (\ref{n1}) must vanish. This condition can be concisely expressed as 
\begin{align}
   \frac{\partial a}{\partial ct} \frac{D \Omega }{D a} + \frac{\partial \vec A}{\partial ct}\cdot \frac{D \Omega}{D \vec A} = 0 \label{gstring2}
\end{align}
Analogous to section \ref{nonrel} in the main text, equation \eqref{gstring2} has the path-dependent solution
\begin{align}
    \Omega \equiv e^{\frac{2g_I}{\hbar c}\int_\mathcal{C}^{(a, \vec A)} (\int_\mathcal M \vec A' \cdot \vec B' \delta a') + (\int_\mathcal M a' \vec B' \cdot \delta \vec A')} = \mathds 1^2 [\mathcal C] \label{solux}
\end{align}
where $\mathcal{C} = \{\big (a'(\lambda), \vec A'(\lambda), t'(\lambda)\big)| \lambda \in [0,1]\}$ defines a trajectory through the configuration space of fields, $\delta a$, $\delta \vec A$ are the field variations along $\mathcal C$ and $\mathds 1[\mathcal C] = e^{\frac{g_I}{\hbar c}\int_\mathcal{C}^{(a, \vec A)} (\int_\mathcal M \vec A' \cdot \vec B' \delta a') + (\int_\mathcal M a' \vec B' \cdot \delta \vec A')}$ is the parallel-transported scale from $[a_0, \vec A_0, t_0] \to [a, \vec A, t]$ along $\mathcal C$. We have set $\Omega[a_0, \vec A_0, t_0] \equiv 1$ in the gauge which satisfies $R[a_0, \vec A_0, t_0] = \tilde R[a_0, \vec A_0, t_0]$, where $\tilde R[a_0, \vec A_0, t_0]$ is the initial gauge-invariant amplitude when $e_I = 0$.\\

The continuity equation (\ref{conjax}) can then be expressed in terms of ordinary field derivatives as
\begin{align}
    \frac{\partial}{\partial t} \bigg (\frac{R^2}{\mathds 1^2[\mathcal C]}\bigg) + \int_\mathcal M \frac{\delta }{\delta a}\bigg( \frac{R^2}{\mathds 1^2[\mathcal C]}(c\frac{\delta S}{\delta a} + g\vec A \cdot \vec B)c\bigg) + \int_\mathcal M\frac{\delta }{\delta \vec A} \cdot \bigg(\frac{R^2}{\mathds 1^2[\mathcal C]}(c\frac{\delta S}{\delta \vec A} + ga\vec B)c\bigg) = 0
\end{align}
Further, the Hamilton-Jacobi equation (\ref{hjax}) can be expressed as
\begin{align}
    \frac{\partial S}{\partial t} + &\frac{1}{2} \int_\mathcal M \bigg \{ \bigg[(c\frac{\delta S}{\delta a} + g\vec A \cdot \vec B )^2 + (\vec \nabla a)^2 + \frac{m^2c^2}{\hbar^2}a^2 \bigg] + \frac{1}{2} \bigg [(c\frac{\delta S}{\delta \vec A} + g a\vec B)^2 + \vec B^2\bigg] \bigg\} \nonumber \\
    & -\frac{\hbar^2c^2}{2}\int_\mathcal M \bigg \{\frac{1}{R}\frac{D^2 R}{Da^2} + \frac{1}{R} \frac{D^2 R}{D \vec{A}^2} \bigg\} = 0 \label{last}
\end{align}
where the quantum potentials for both the axion field and the gauge field have been concisely expressed using the Weyl covariant field derivatives defined by the equations \eqref{fcove1}, \eqref{fcove2}. The covariant field derivatives here act on $R$, which has Weyl weight $\omega = 1$. 

\subsection{Gauge invariance} \label{axio3}
Let us check the gauge invariance of our formulation. We note that the Weyl gauge allows residual \textit{spatial} gauge transformations $A_\mu \to A_\mu - \partial_\mu \lambda(\vec x)$. Upon an infinitesimal spatial gauge transformation (small $\partial_\mu\lambda$), it is straightforward to check that the Schrödinger equation (\ref{axmain}) in the main text remains invariant if the quantum state transforms as $\Psi' \to \Psi e^{-i\frac{g_C}{\hbar c} \int_\mathcal{M} a \vec \nabla \lambda \cdot \vec B }$. It follows that that the phase changes from $S \to S' = S - \frac{g}{c}\int_\mathcal{M} a\vec B \cdot \vec \nabla \lambda $ and the amplitude changes from $R \to R' = Re^{\frac{g_I}{\hbar c}\int_\mathcal{M} a\vec B \cdot \vec \nabla \lambda}$. It is straightforward to check that the guidance equations (\ref{guy1}), (\ref{guy2}) in the main text thereby remain invariant. On the other hand, under the infinitesimal gauge transformation, the weight factor changes from
\begin{align}
    \Omega \to \Omega' = e^{\frac{2g_I}{\hbar c}\int_\mathcal{M} a\vec B_0 \cdot \vec \nabla \lambda_0} \times e^{\frac{2g_I}{\hbar c}\int_{(a_0, \vec A_0 + \vec \nabla \lambda_0)}^{(a, \vec A + \vec \nabla \lambda)} (\int_\mathcal M \vec A'\cdot \vec B' \delta a') + (\int_\mathcal M a' \vec B' \cdot \delta \vec A') } = e^{\frac{2g_I}{\hbar c} \int_\mathcal M a \vec B \cdot \vec \nabla \lambda }
\end{align}
where the initial scale at $[a_0, \vec A_0, t_0]$ has changed from $1 \to 1\cdot e^{\frac{2g_I}{\hbar c}\int_\mathcal{M} a\vec B_0 \cdot \vec \nabla \lambda_0}$. Therefore, the current density $ R^2/\Omega = R'^2/\Omega'$ is gauge invariant. Lastly, it is clear from equation (\ref{last}) that the quantum potentials for the axion and the gauge field, defined using the Weyl covariant derivatives, are gauge invariant.

\end{document}